\documentclass[10pt,twocolumn]{revtex4-2}

\usepackage{graphicx}
\usepackage[percent]{overpic}
\usepackage{amsmath,amssymb,bm}
\usepackage{booktabs}
\usepackage{xcolor}
\usepackage[colorlinks=true,citecolor=blue!55!black,linkcolor=blue!55!black,urlcolor=blue!55!black]{hyperref}

\graphicspath{{figures/}}
\newcommand{\vq}{\bm q}
\newcommand{\vrnew}{\bm r}
\newcommand{\vu}{\bm u}

\newcommand{\avg}[1]{\left\langle #1\right\rangle}
\newcommand{\dd}{\mathrm d}
\newcommand{\ii}{\mathrm i}
\newcommand{\kb}{k_{\mathrm B}}
\newcommand{\Cu}{C^u}

\newcommand{\Hcov}{H^{\mathrm{cov}}}
\newcommand{\SK}[1]{\textcolor{black}{{#1}}}

\begin{document}

\title{Enhanced Long-Wavelength Fluctuations and Interaction-Stress Correlations in Active Crystals}

\author{Antik Bhattacharya}
\email{abhattacharya@tifrh.res.in}
\author{Smarajit Karmakar}
\email{smarajit@tifrh.res.in}
\affiliation{Tata Institute of Fundamental Research Hyderabad, 36/P, Gopanpally Village, Serilingampally Mandal, Ranga Reddy District, Hyderabad, Telangana 500046, India}

\begin{abstract}
\SK{Activity can enhance the long-wavelength density fluctuations of a solid beyond the equilibrium fluctuations predicted by the Mermin-Wagner-Hohenberg theorem, but the microscopic origin of this enhancement remains unclear. We study a two-dimensional active Brownian crystal whose longitudinal and transverse displacement covariances scale as $\Cu_\lambda(q)\sim q^{-3}$ as compared to the passive counterpart with  $q^{-2}$ law. This spectrum predicts an MSD-plateau divergence proportional to $L$ in two dimensions, logarithmic in three dimensions, and finite in four dimensions, in close correspondence with recent results reported in Dey \textit{et al.} [\href{https://doi.org/10.1038/s41467-025-61366-0}{\textit{Nat. Commun.} \textbf{16}, 5498 (2025)}]. Inverse of the Covariance gives $\Hcov_\lambda\sim |q|^3$ with the corresponding dispersion $\omega_{\mathrm{cov},\lambda}\sim q^{3/2}$, whereas direct mechanical response gives $K_{\mathrm{resp},\lambda}\sim q^2$ and acoustic modes with $\omega_{0,\lambda}\sim q$. The nonlinear dispersion is therefore covariance-defined rather than mechanical. The mobility and integrated active-force correlations remain non-singular, although a finite-persistence crossover cannot be ruled out completely with the existing data. Microscopically, the longitudinal and transverse Irving--Kirkwood interaction-stress spectra scale as $q^{-1}$. The transverse stress-stress correlation connects well with the anomalous displacement field, indicating that activity anomalously populates ordinary acoustic modes, while the conservative interaction network transmits the resulting fluctuations as long-ranged stress.  
A three-dimensional active FCC crystal exhibits the same paired low-$q$ trends.}
\end{abstract}

\maketitle

\section{Introduction}

\SK{Long-wavelength modes place a fundamental restriction on
positional order in low-dimensional (2D) solids.  In solids, these long-wavelength modes correspond to acoustic phonons, whose frequency vanishes linearly with the wave vector. The mean fluctuation of a particle $\vec{u}(\vec{r})$ at position $\vec{r}$, receives contributions from phonon modes of all wavelengths, and in 2D, the mean-squared displacement fluctuation $C^u_\lambda(q) = \langle |u(q)|^2\rangle$ can be written within the harmonic approximation as 
\begin{equation}
 C^u_\lambda(q)=\frac{\kb T}{\omega(q)^2},
 \label{eq:passiveCu}
\end{equation}
where $\omega(q)$ is the frequency of the mode with wave vector $q$, $\kb$ is the Boltzmann constant and $T$ is the bath temperature. These modes fall into two classes: transverse (subscript T) and longitudinal (subscript L). Translational invariance and short-ranged elasticity enforce $\omega(q)$, which is also connected to the mechanical kernel $\omega_\lambda^2(q) \propto K_\lambda(q) \simeq M_\lambda q^2+O(q^4)$, where $\lambda=L,T$ denotes the
longitudinal or transverse polarisation and $M_\lambda$ is the corresponding elastic modulus. This directly produces the logarithmic Mermin-Wagner-Hohenberg (MWH) displacement divergence in two dimensions
\cite{Hohenberg1967,MerminWagner1966,Mermin1968,Jancovici1967,ImryGunther1971}.  The same collective fluctuations are relevant not only to crystalline order but also to the dynamics of two-dimensional amorphous solids, colloidal glasses, and liquids
\cite{Shiba2016,Flenner2015,Vivek2017,Illing2017,Li2019,Mazoyer2009}. In three dimensions (3D), the phase-space factor renders the $q^{-2}$ spectrum infrared convergent, so the corresponding displacement plateau remains finite.}

\SK{Activity changes this balance because energy is injected locally and detailed balance no longer constrains the occupation of elastic modes \cite{Ramaswamy2010,Marchetti2013,Bechinger2016,CatesTailleur2015,Fodor2016}. Dense active systems also display freezing, jamming, glassy arrest, and rigidity at non-equilibrium conditions \cite{Fily2014,Henkes2011,BerthierKurchan2013,
Berthier2014,Berthier2019,Paul2023,Bi2015,Henkes2020}.  Previous studies of active crystals and glasses in the presence of run-and-tumble particles (RTPs) or active Brownian particles (ABPs) found that activity amplifies long-wavelength modes far beyond their passive value \cite{Dey2025}.  The plateau of the mean-squared displacement fluctuations grew as a power of system size in two
dimensions, logarithmically in three dimensions, and remained finite in four dimensions (4D).  By measuring particle displacements relative to their neighbouring particles and calculating the corresponding correlation functions—commonly referred to as cage-relative observables \cite{}—the size dependence was eliminated. This demonstrates that the observed anomaly originates from long-wavelength collective modes, rather than from local cage rearrangements.  An effective dynamical matrix \cite{Dey2025} constructed by inverting the displacement covariance matrix (see the Model and Observables section for the definition) showed enhanced low-frequency spectral weight and an apparent nonlinear dispersion
$\omega_{\mathrm{eff}}\sim q^\alpha$, with effective exponents exceeding the passive value $\alpha=1$ and, for several branches, approaching $3/2$. Related studies connect persistent or correlated driving to anomalous long-wavelength fluctuations \cite{Galliano2023,Ikeda2023,Kuroda2023}.}

\SK{Those observations exposed a central unresolved issue.  At equilibrium, fluctuation--dissipation theory makes the inverse covariance proportional to the mechanical Hessian, so a covariance-derived dispersion can be identified with a phonon dispersion \cite{Henkes2012,Still2013}.  This identification is generally not valid in an active steady state.  The inverse covariance describes
the probability with which a deformation is sampled, whereas the response kernel measures the force required to impose that deformation.  Therefore, the previously observed nonlinear dispersion could signify a genuinely softened mechanical operator, an anomalous scale-dependent occupation of otherwise ordinary acoustic modes, or a combination of the two.  Determining which interpretation is correct is necessary both to identify the microscopic origin of the fluctuations and to assign a physical meaning to the shift of the marginal dimension \cite{Loi2008,Szamel2014,Maggi2015,Cugliandolo2011}.}

\SK{Several distinct mechanisms can supply the additional infrared weight.  The elastic kernel itself could become nonlocal, with $K\sim |q|^3$; a local $q^2$ kernel could be driven by a scale-dependent mode temperature; finite persistence could create a crossover with an apparent local $q^{-3}$ slope; the mobility or integrated active-force intensity could become singular; or activity could generate a collective long-ranged stress field.  These possibilities lead to the same or similar equal-time displacement spectra but make different predictions for response, temporal correlations, and microscopic force balance \cite{Takatori2014,Solon2015,Omar2020,Cates2008,Maitra2019,Scheibner2020,Banerjee2021,Ronceray2015}.  Equal-time covariance alone will not be able to distinguish amongst these possibilities \cite{DalCengio2019,Mandal2017}.}

\SK{Here we make this distinction for an active Brownian crystal by combining the displacement covariance with paired static forcing, frequency-dependent response, mode relaxation, active- and residual-force correlations, and Irving-Kirkwood interaction stress fluctuations. We first establish displacement fluctuation $\Cu_\lambda(q)\sim q^{-3}$ for low $q$ modes. This single exponent organizes the earlier results of dimension dependence of the mean squared displacement plateau with system size, as integrating the spectrum gives mean squared displacement in two, three, and four dimensions as $W_2\sim L$, $W_3\sim\ln L$, and a finite $W_4$, respectively, while inversion of the covariance matrix will give the corresponding eigenvalues to scale as $\sim |q|^3$, leading to $\omega_{\mathrm{cov},\lambda}\sim q^{3/2}$. Direct mechanical measurements, however, give $K_{\mathrm{resp},\lambda}\sim q^2$. The nonlinear law is therefore a covariance-defined dispersion that encodes anomalous mode occupation, not a cubic mechanical phonon operator. We further test the remaining routes to this anomalous occupation. The mobility, propulsion persistence, and integrated residual-force intensity remain nonsingular. By contrast, both longitudinal and transverse Irving-Kirkwood interaction-stress spectra grow approximately as $q^{-1}$ for small $q$. A three-dimensional FCC calculation exhibits the same paired low-$q$ trends in displacement and stress correlation. This suggests that long range stress correlation to be primary reason behind the results reported in \cite{Dey2025}, as elaborated in the later part of this article.} 

\SK{This article is structured as follows. The model under investigation is first introduced, followed by a description of the methods used to compute the response, force correlations, and microscopic stress correlations. The results concerning enhanced displacement fluctuations and the influence of nonlocal elastic softening are then presented. The discussion then addresses the potential for mode-dependent effective temperature and the possible contribution of nonlocal mobility to the observed enhanced displacement fluctuations. The analysis further explores the possibility of wave vector-dependent active force correlations as an underlying cause of these anomalous displacement fluctuations. The article concludes that long-range stress correlations may be responsible for the observed phenomena.}

\section{Model and observables}

\subsection{Active Brownian crystal}

\SK{We simulate $N$ identical particles in a two-dimensional periodic box,
initialized on a triangular lattice.  The equations of motion for their underdamped active Brownian motion are \cite{Fily2014,Palacci2013,Bechinger2016}
\begin{align}
 m\ddot{\vrnew}_i={}&-\gamma\dot{\vrnew}_i+\sum_{j\ne i}\bm F_{ij}
 +f_0\bm n_i+\bm\xi_i,                                         \label{eq:eom}\\
 \dot\theta_i={}&\sqrt{2D_r}\,\eta_i,\qquad
 \bm n_i=(\cos\theta_i,\sin\theta_i),                          \label{eq:orientation}
\end{align}
where $m$, $\gamma$, $f_0$, and $D_r$ are the particle mass, dissipative force strength, active force strength, and rotational diffusion constant, respectively. $\mathbf r_i$ and $\theta_i$ are the particle positions and the active-force orientation angle, and $\mathbf F_{ij}$ is the conservative pair force. The translational and rotational noises are independent Gaussian white noises, with
\begin{align}
 \avg{\xi_{i\alpha}(t)\xi_{j\beta}(t')}&=
 2\gamma\kb T\,\delta_{ij}\delta_{\alpha\beta}\delta(t-t'),\\
 \avg{\eta_{i}(t)\eta_{j}(t')}&=
 \delta_{ij}\delta(t-t').
\end{align}
where $k_{\mathrm B}$ is the Boltzmann constant, and $T$ is translational temperature of the system. The conservative force follows from the Lennard--Jones potential
\begin{equation}
 U(r)=4\epsilon\left[\left(\frac{\sigma}{r}\right)^{12}
 -\left(\frac{\sigma}{r}\right)^6\right],
\end{equation}
truncated at $r_c=2.5\sigma$, where $\epsilon$ determines the strength of LJ potential and $\sigma$ is diameter of each particle. We use reduced units
$m=\sigma=\epsilon=1.0$.  The baseline state has number density $\rho=1.0$,
$\gamma=5.0$, $\kb T=0.01$, $f_0=2.0$, $D_r=0.1$, and time step
$\Delta t=10^{-3}$.  The principal response and stress measurements use a
$16\times16$ lattice.  Each trajectory is equilibrated for $10^5$ steps and
sampled from $32$ statistically independent production runs.}

\SK{Writing $\vu_i=\vrnew_i^{\,u}-\bm R_i$, where $\bm R_i$ is the mean reference position of the particles, for the unwrapped displacement from the reference lattice, we remove the center-of-mass mode and define
\begin{equation}
 \vu(\vq,t)=\frac{1}{\sqrt N}\sum_i\vu_i(t)e^{-\ii\vq\cdot\bm R_i}.
\end{equation}
With $\hat{\bm e}_L=\hat{\vq}$ and
$\hat{\bm e}_T=(-\hat q_y,\hat q_x)$,
\begin{equation}
 u_\lambda(\vq,t)=\hat{\bm e}_\lambda\cdot\vu(\vq,t),\qquad
 \Cu_\lambda(q)=\avg{|u_\lambda(\vq)|^2}.
 \label{eq:CuDef}
\end{equation}
Equivalent wave vectors and independent trajectories are averaged before
radial binning.}

\subsection{Response, force correlations, and mobility}

\SK{We measure the static susceptibility using paired sinusoidal body forces
$\pm A\hat{\bm e}_\lambda\cos(\vq\cdot\bm R_i)$ applied to trajectories
with common noise.  The odd response eliminates the unforced background,
For each probed wave vector, we record the cosine amplitude
of the center-of-mass-subtracted displacement,
\begin{equation}
 U_\lambda^c(\vq,t)
 =\frac{2}{N}\sum_i
 \left[\hat{\bm e}_\lambda\cdot\widetilde{\vu}_i(t)\right]
 \cos(\vq\cdot\bm R_i),
 \label{eq:cosineResponseAmplitude}
\end{equation}
where $\widetilde{\vu}_i=\vu_i-N^{-1}\sum_j\vu_j$. The selected nonzero lattice modes satisfy
$\sum_i\cos^2(\vq\cdot\bm R_i)=N/2$, so the factor $2/N$ recovers the amplitude of a cosine
deformation.}

\SK{The static susceptibility and response stiffness are
\begin{equation}
 \chi_\lambda(q)
 =\frac{\avg{U_\lambda^c(\vq)}_+
             -\avg{U_\lambda^c(\vq)}_-}{2A},
 \qquad
 K_{\mathrm{resp},\lambda}(q)=\chi_\lambda^{-1}(q),
 \label{eq:responseDef}
\end{equation}
where $A$ is the applied force amplitude per mode, the subscripts distinguish the positive and negative probe trajectories, and the brackets denote steady-state time and ensemble averages.}

\SK{The cosine amplitude is related to the Fourier mode defined above by
\begin{equation}
 U_\lambda^c(\vq,t)
 =\frac{2}{\sqrt N}\operatorname{Re}u_\lambda(\vq,t).
\end{equation}
Thus no additional system-size factor is required in $\chi_\lambda$ or $K_{\mathrm{resp},\lambda}$.
The sine quadrature is obtained by replacing the cosine with a sine in Eq.~\eqref{eq:cosineResponseAmplitude}. Probe amplitudes $A=5\times10^{-4},10^{-3},2\times10^{-3}$ give the same
susceptibility.  The quadrature response is below $10^{-3}$ of the in-phase component.}

\SK{For the projected propulsion force
\begin{equation}
 f^a_\lambda(\vq,t)=\frac{f_0}{\sqrt N}\sum_i
 \hat{\bm e}_\lambda\cdot\bm n_i(t)e^{-\ii\vq\cdot\bm R_i},
\end{equation}
we compute the active force-force correlation time and integrated intensities as
\begin{align}
 C^f_\lambda(q,t)&=\avg{f^a_\lambda(\vq,t_0+t)
 [f^a_\lambda(\vq,t_0)]^*},                                      \label{eq:Cf}\\
 \tau^f_\lambda(q)&=\frac{1}{C^f_\lambda(q,0)}
 \int_0^\infty C^f_\lambda(q,t)\,\dd t,                      \label{eq:tauf}\\
 \mathcal D^f_\lambda(q)&=\int_0^\infty C^f_\lambda(q,t)\,\dd t.
 \label{eq:Df}
\end{align}
Similarly, we will calculate the residual force-force correlation time and integrated intensities. By fitting the Dynamic step-response, we can independently compute the mode stiffness and damping
$\zeta_\lambda(q)$, and mobility $\mathcal M_\lambda(q)=1/\zeta_\lambda(q)$.}

\subsection{Microscopic stress}
\SK{The Irving--Kirkwood tensor is a microscopic interaction stress tensor, while Hooke's law is a continuum constitutive approximation. We test whether the connected interaction-stress fluctuations are dominated by the linear elastic contribution at long wavelengths. The Fourier-space Irving--Kirkwood interaction stress tensor is evaluated directly from LJ pair forces \cite{IrvingKirkwood1950,Hardy1982,AdmalTadmor2010}:
\begin{equation}
 \sigma^{\mathrm{int}}_{\alpha\beta}(\vq)=
 -\frac{1}{V}\sum_{i<j}r_{ij,\alpha}F_{ij,\beta}
 e^{-\ii\vq\cdot\vrnew^{\mathrm{mid}}_{ij}}
 \operatorname{sinc}\!\left(\frac{\vq\cdot\vrnew_{ij}}{2}\right).
 \label{eq:IK}
\end{equation}
where $V$ is the area in $2D$ and volume in $3D$, $\operatorname{sinc}x=\frac{\sin x}{x}$, with value $1$ at zero. The minimum-image convention is used, and the sinc factor is the exact Fourier
transform of a straight force bond, and $\vrnew^{\mathrm{mid}}_{ij}=\frac{{\bf r}_i+{\bf r}_j}{2}$.  Next, we project
$\sigma_{LL}^{\mathrm{int}}({\bf q})=\hat{e}_{L,\alpha}\cdot\bm\sigma_{\alpha \beta}^{\mathrm{int}}\cdot\hat{e}_{L, \beta}$ and
$\sigma_{TL}^{\mathrm{int}}({\bf q})=\hat{e}_{T,\alpha}\cdot\bm\sigma_{\alpha \beta}^{\mathrm{int}}\cdot\hat{e}_{L, \beta}$ and define stress variances as $S_{\text{stress},L}(q)=\avg{|\delta\sigma_{LL}^{\text{int}}({\bf q})|^2}$ and $S_{\text{stress},T}(q)=\avg{|\delta\sigma_{TL}^{\text{int}}({\bf q})|^2}$. Longitudinal fluctuations couple to compression, density variations, bulk modulus, pressure, and transverse fluctuations primarily represent shear.}

\section{Results}

\subsection{Enhanced displacement fluctuations}

\begin{figure*}[t]
 \centering
 \begin{overpic}[width=0.48\linewidth]{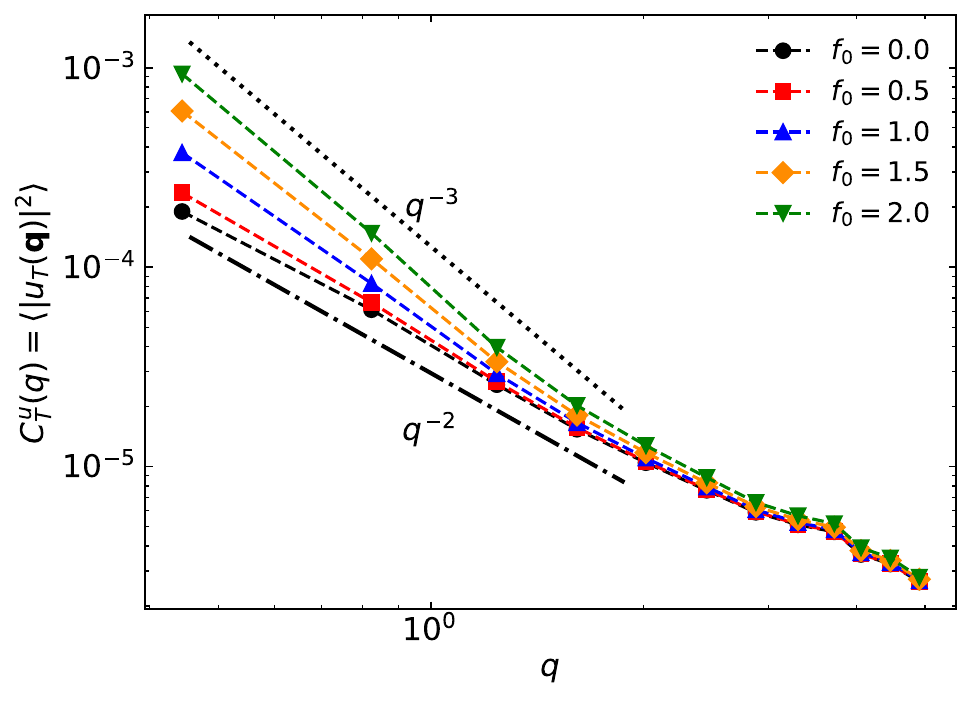}
    \put(17,13){\large\bfseries (a)}
\end{overpic}
\begin{overpic}[width=0.48\linewidth]{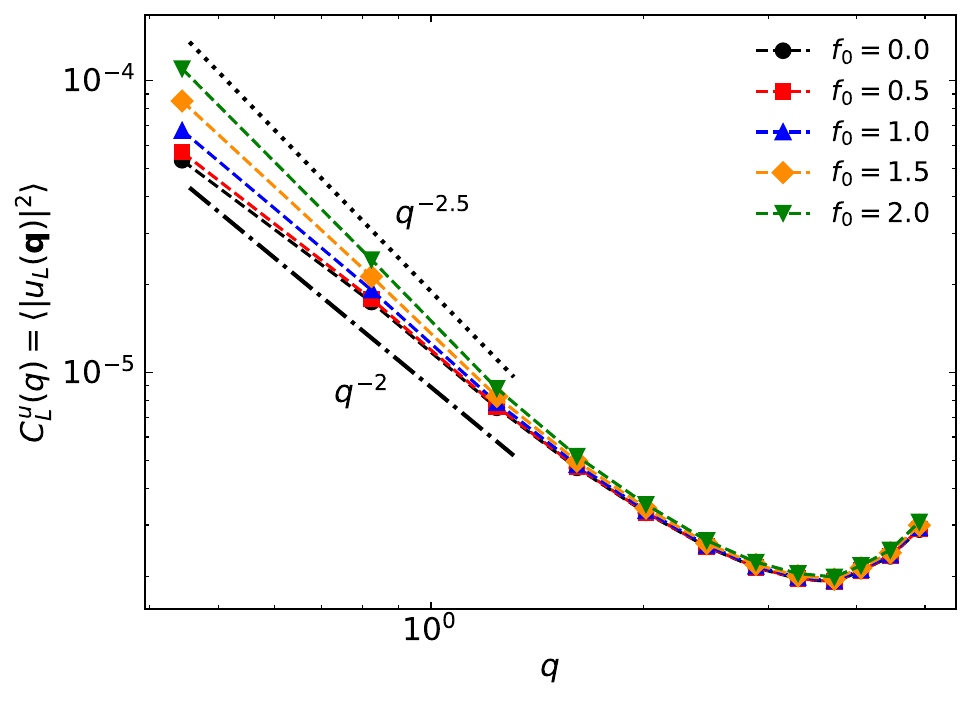}
    \put(17,13){\large\bfseries (b)}
\end{overpic}
 \caption{Activity-enhanced displacement fluctuations in the two-dimensional crystal.
(a) Transverse displacement covariance $C_T^u(q)=\langle |u_T(\vq)|^2\rangle$ and
(b) longitudinal displacement covariance $C_L^u(q)=\langle |u_L(\vq)|^2\rangle$ for the indicated propulsion strengths $f_0$. The passive spectra are consistent with the conventional elastic scaling $C_\lambda^u(q)\sim q^{-2}$. Increasing activity selectively
enhances the long-wavelength modes. At low-$q$, the
transverse spectrum approaches the $q^{-3}$ guide, whereas the longitudinal
spectrum displays a weaker effective dependence, approximately $q^{-2.5}$.}
 \label{fig:Cu}
\end{figure*}

\SK{The passive crystal follows the elastic $q^{-2}$ law, whereas activity progressively enhances the long-wavelength transverse and longitudinal modes to $q^{-3}$ as shown in Figure~\ref{fig:Cu} (a) and (b), respectively.  For the active state, both polarizations are consistent over the resolved interval with
\begin{equation}
 \Cu_\lambda(q)\simeq A_\lambda q^{-3}.
 \label{eq:qminus3}
\end{equation}
This equal-time spectrum alone does not uniquely determine its microscopic origin, where $A_\lambda$ is the amplitude of the low-wave-number displacement spectrum. In a nonequilibrium system, both the response and the statistics of the active forcing can be scale-dependent. A useful diagnostic decomposition can be
\begin{equation}
  C^u_\lambda(q)
  =
  \frac{T_{\mathrm{eff},\lambda}(q)}
       {K_{\mathrm{resp},\lambda}(q)},
  \label{eq:diagnostic-decomposition}
\end{equation}
where $K_{\mathrm{resp},\lambda}(q)$ is the stiffness obtained from an independent mechanical-response measurement and $T_{\mathrm{eff},\lambda}(q)$ parametrizes the strength of the nonequilibrium fluctuations (the so-called effective temperature). Conventional long-wavelength elasticity gives $K_{\mathrm{resp},\lambda}(q)\sim q^2$.  Hence the extra factor $q^{-1}$ in Eq.~\eqref{eq:qminus3} could well originate from an anomalous response, anomalous active fluctuations, or a crossover between different asymptotic regimes.}

\vskip +0.1in
\noindent{\bf Central scaling theorem and the lower critical dimension:} \SK{For a general infrared spectrum $C^u_\lambda(q)\simeq A_\lambda(\widehat{\vq})q^{-\psi}$. Let $\rho=N/V$ be the number density and $\delta u_\lambda=u_\lambda-\langle u_\lambda\rangle$. Throughout, $C_\lambda^u(q)=\langle|\delta u_\lambda(\bm q)|^2\rangle$ denotes the connected displacement covariance after removing translation. For a complete orthonormal set of lattice Fourier modes, the mean squared displacement from the reference position of the crystal is
\begin{align}
 W_d(L)&\equiv\frac{1}{N}\sum_i
       \left\langle|\delta\bm u_i|^2\right\rangle
       =\frac{1}{N}\sum_{q\ne0,\lambda}C_\lambda^u(q),\\
 &\simeq
       \frac{1}{\rho}\sum_\lambda
       \int_{2\pi/L}^{\Lambda}
       \frac{d^d q}{(2\pi)^d}\,C_\lambda^u(q).
 \label{eq:W_def}
\end{align}
where $\Lambda$ is a microscopic cutoff.  After angular averaging, define
\begin{equation}
 \mathcal A_d=\frac{1}{S_{d-1}}\int\dd\Omega
 \sum_\lambda A_\lambda(\widehat{\vq}),
 \qquad
 S_{d-1}=\frac{2\pi^{d/2}}{\Gamma(d/2)}.
\end{equation}
where $d$ is the spatial dimension. $\mathrm d \Omega$ is solid-angle measure, $S_{d-1}$ is the area of the unit sphere in $d$ dimensions, and $\Gamma$ is the the Euler gamma function. $\mathcal A_d$ is the angular average of the sum over all polarization amplitudes. Then
\begin{equation}
 W_d(L)=\frac{S_{d-1}\mathcal A_d}{(2\pi)^d \rho}
 \int_{2\pi/L}^{\Lambda}q^{d-1-\psi}\dd q.
 \label{eq:master_integral}
\end{equation}
For $d\neq\psi$,
\begin{equation}
 W_d(L)=\frac{S_{d-1}\mathcal A_d}{(2\pi)^d \rho}
 \frac{\Lambda^{d-\psi}-(2\pi/L)^{d-\psi}}{d-\psi},
\end{equation}
while for $d=\psi$,
\begin{equation}
 W(L)=\frac{S_{d-1}\mathcal A_d}{(2\pi)^d \rho}
 \ln\left(\frac{\Lambda L}{2\pi}\right).
\end{equation}
Therefore, at a constant number density
\begin{equation}
 W_d(L)\sim
 \begin{cases}
 L^{\psi-d},&d<\psi,\\
 \ln L,&d=\psi,\\
 \mathrm{constant},&d>\psi.
 \end{cases}
 \label{eq:general_finite_size}
\end{equation}
The marginal dimension is consequently \cite{Hohenberg1967,Ikeda2023}
\begin{equation}
 d_\ell=\psi.
 \label{eq:dl_sigma}
\end{equation}
If the covariance-defined dispersion is $\omega_{\rm cov}\sim q^\alpha$, then
$C^u\propto\omega_{\rm cov}^{-2}$ gives $\psi=2\alpha$.  Hence the same result
may be written
\begin{equation}
 d_\ell=2\alpha,\qquad
 W_d(L)\sim L^{2\alpha-d}\quad(d<2\alpha).
 \label{eq:dl_alpha}
\end{equation}}

\vskip +0.05in
\noindent{\bf Passive and active prediction:} \SK{The plateau of the time-dependent MSD, $\Delta(t)=\avg{|\vu(t)-\vu(0)|^2}$, satisfies $\text{MSD}_{\rm pl}=2W$ when long-time displacement correlations have decayed and the centre-of-mass mode is removed. For passive solid with $\psi=2$, $d_\ell=2$, in two dimensions,
\begin{equation}
 W^{\rm p}_{2D}(L)
 =\frac{\mathcal P_2}{2\pi \rho}
 \ln\left(\frac{\Lambda L}{2\pi}\right),
 \qquad
 \text{MSD}_{\rm pl}^{\rm p,2D}\sim\ln L,
 \label{eq:passive_2d}
\end{equation}
where $\mathcal P_2$ is the sum of the passive longitudinal and transverse
amplitudes.  In three dimensions,
\begin{equation}
 W^{\rm p}_{3D}(L)
 =\frac{\mathcal P_3}{2\pi^2 \rho}
 \left(\Lambda-\frac{2\pi}{L}\right),
 \qquad
 \text{MSD}_{\rm pl}^{\rm p,3D}\to\mathrm{constant}.
 \label{eq:passive_3d}
\end{equation}
where, $\mathcal P_3 = P_L +2 P_T$, where $P_\lambda$ is the coefficient of $q^{-2}$in the covariance convention. For active solid with $\psi=3$, $d_\ell=3$, the measured active spectrum, Eq.~\eqref{eq:master_integral} yields in two dimensions
\begin{equation}
 W^{\rm a}_{2D}(L)
 =\frac{\mathcal A_2}{2\pi \rho}
 \left(\frac{L}{2\pi}-\frac{1}{\Lambda}\right),
 \qquad
 \text{MSD}_{\rm pl}^{\rm a,2D}\sim L.
 \label{eq:active_2d}
\end{equation}
In three dimensions,
\begin{equation}
 W^{\rm a}_{3D}(L)
 =\frac{\mathcal A_3}{2\pi^2 \rho}
 \ln\left(\frac{\Lambda L}{2\pi}\right),
 \qquad
 \text{MSD}_{\rm pl}^{\rm a,3D}\sim\ln L.
 \label{eq:active_3d}
\end{equation}
In four dimensions,
\begin{equation}
 W^{\rm a}_{4D}(L)
 =\frac{\mathcal A_4}{8\pi^2 \rho}
 \left(\Lambda-\frac{2\pi}{L}\right),
 \qquad
 \text{MSD}_{\rm pl}^{\rm a,4D}\to\mathrm{constant}.
 \label{eq:active_4d}
\end{equation}
The observed change in the critical dimensions therefore results solely from a single exponent change, $\psi:2\to3$, or equivalently $\alpha:1\to3/2$. Dey \emph{et al.}\ reported that the passive MSD plateau grows logarithmically with system size $L$ in two dimensions and remains finite in three dimensions. In contrast, active systems exhibit power-law growth in two dimensions, logarithmic growth in three dimensions, and no divergence in four dimensions \cite{Dey2025}. The present results provide a compact asymptotic organizing principle for these observations. The passive control establishes $\psi_{\rm p}=2$, whereas the compensated active spectrum supports $\psi_{\rm a}=3$ over the resolved wave-vector window, corresponding to a covariance-derived dispersion exponent $\alpha=3/2$. Equation~\eqref{eq:dl_alpha} then predicts precisely the dimensional sequence reported previously. Deviations of the effective two-dimensional finite-size exponent from unity, or branch-dependent effective dispersion exponents from $3/2$, may arise from finite-window corrections to the asymptotic scaling $C_\lambda^u(q)\sim q^{-3}$. Further studies are needed to ascertain the validity of this argument.}

\SK{The physical origin of this anomalous displacement spectrum, however, remains unclear. Possible mechanisms include emergent nonlocal elastic softening, a scale-dependent active mode temperature, nonlocal mobility, wave-vector-dependent active-force correlations, and an emergent long-ranged active-stress field. We test these hypotheses systematically in the following sections.}

\begin{figure*}[t]
 \centering
 \begin{overpic}[width=0.48\linewidth]{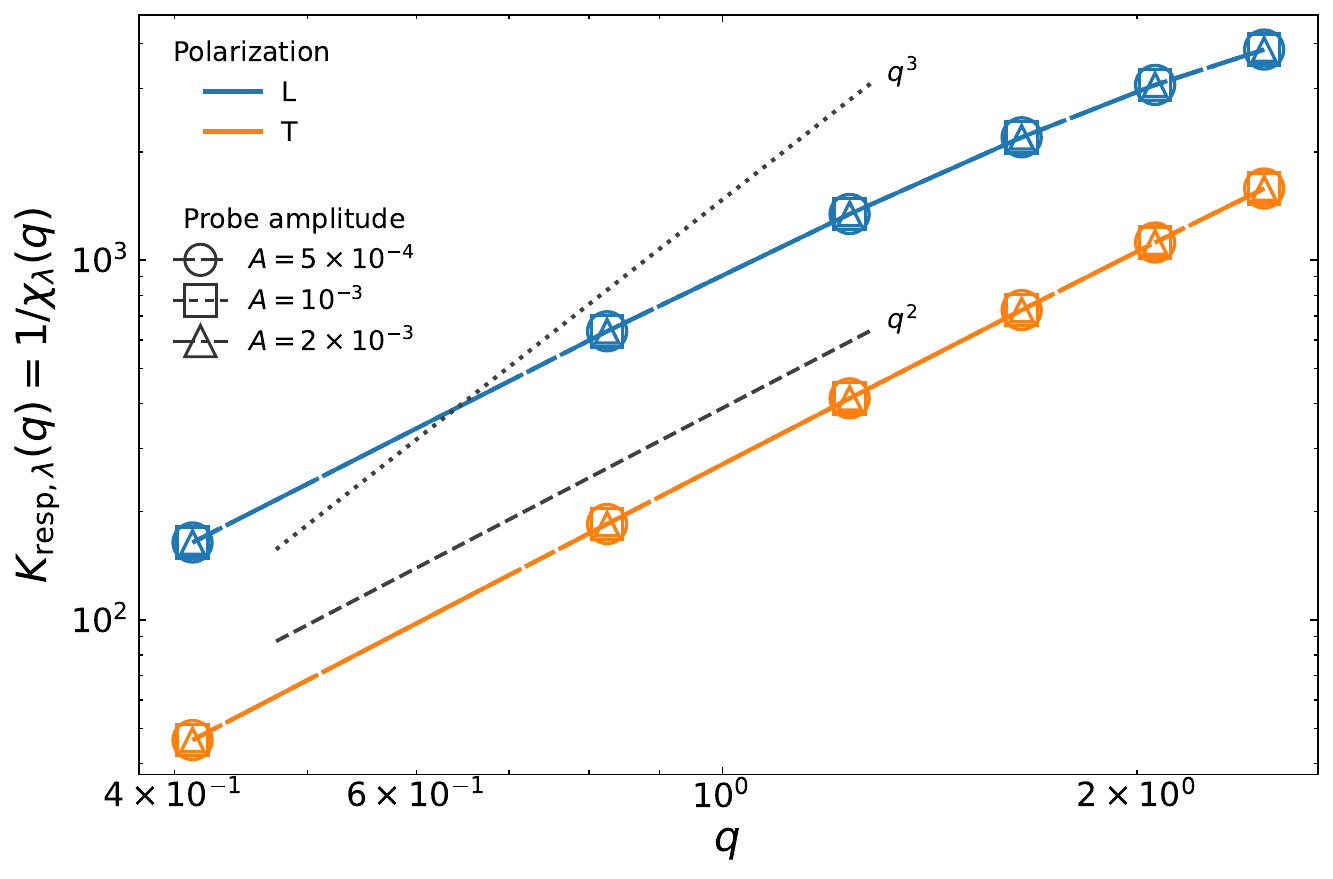}
    \put(88,13){\large\bfseries (a)}
\end{overpic}
\begin{overpic}[width=0.48\linewidth]{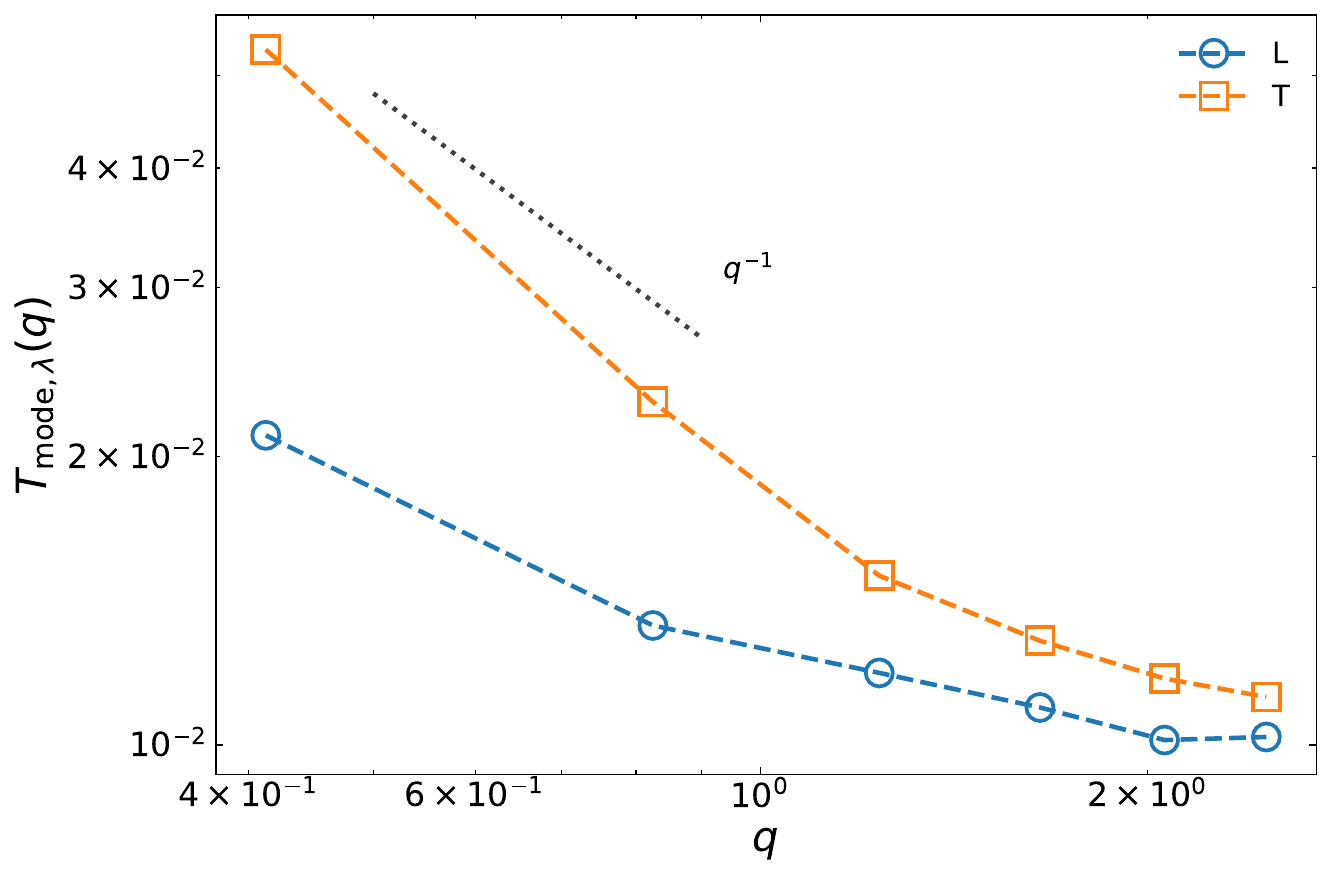}
    \put(20,13){\large\bfseries (b)}
\end{overpic}
 \caption{(a) Static response stiffness
$K_{\mathrm{resp},\lambda}(q)=\chi_\lambda^{-1}(q)$ for the longitudinal and transverse
polarizations. Different open symbols denote the probe amplitudes
$A=5\times10^{-4}$, $10^{-3}$, and $2\times10^{-3}$ and $f_0=2.0$. Their collapse
demonstrates that the probe lies within the linear-response regime. Both
polarizations exhibit the ordinary elastic scaling
$K_{\mathrm{resp},\lambda}(q)\sim q^2$, rather than the $|q|^3$ behavior inferred from
covariance inversion. The dashed and dotted lines indicate $q^2$ and $q^3$
guides, respectively. (b) Equal-time mode-dependent fluctuation--response ratio
$T_{\mathrm{mode},\lambda}(q)=K_{\mathrm{resp},\lambda}(q)\Cu_\lambda(q)$ for the longitudinal and transverse polarizations for $f_0=2.0$.  Both branches increase toward small $q$, demonstrating that
activity enhances the long-wavelength modes.  The transverse branch is approximately compatible with the indicated $q^{-1}$ guide, whereas the longitudinal branch
exhibits a weaker increase. 
}
 \label{fig:modeT}
\end{figure*}

\subsection{Emergent nonlocal elastic softening}
\label{sec:nonlocal-softening}

The first possibility is that activity modifies the mechanical restoring
operator itself. In harmonic equilibrium, after removing the translational
zero modes, fluctuation--dissipation theory gives
\begin{equation}
\bm C=k_{\rm B}T\bm H^{-1},
\end{equation}
so that displacement-displacement covariance inversion recovers the harmonic Hessian
\cite{Henkes2012,Still2013,PhysRevLett.118.098001}. Motivated by this equilibrium identity, we define the covariance operator for each eigenmode as
\begin{equation}
\Hcov_\lambda(q)
=T_0[\Cu_\lambda(q)]^{-1},
\label{eq}
\end{equation}
where $T_0$ is an arbitrary reference energy that fixes the overall units but
not affect the wave-number dependence. The measured active covariance $\Cu_\lambda(q)\simeq A_\lambda |q|^{-3}$ therefore gives
\begin{align}
 \Hcov_\lambda(q)&\simeq\frac{T_0}{A_\lambda}|q|^3,
 \nonumber\\
 \omega_{\mathrm{cov},\lambda}(q)
 &=\sqrt{\Hcov_\lambda(q)/m}\sim q^{3/2}.
 \label{eq:omegaCov}
\end{align}
This construction is exact as a representation of the equal-time covariance,
but it does not yet identify a mechanical phonon frequency. Since $H_{\lambda}^{\text{cov}}(q) \propto |q|^3$, the stationary distribution can be presented schematically as
\begin{align}
 \mathcal P[\vu]&\propto e^{-\mathcal S_{\mathrm{cov}}[\vu]},\nonumber\\
 \mathcal S_{\mathrm{cov}}[\vu]
 &=\frac{1}{2T_0}\sum_{\vq,\lambda}
 \Hcov_\lambda(q)|u_\lambda(\vq)|^2\nonumber\\
 &\simeq\frac{1}{2}\sum_{\vq,\lambda}
 \frac{|q|^3}{A_\lambda}|u_\lambda(\vq)|^2.
 \label{eq:covFunctional}
\end{align}
In real space, $|q|^3$ corresponds to the fractional operator $(-\nabla^2)^{3/2}$,
Eq.~\eqref{eq:covFunctional}, spatially nonlocal in the probability distribution of the active displacement field. The important question is whether the same operator appears in the force required to deform the crystal.


\SK{If the mechanical kernel were genuinely cubic, it could be written as
\begin{equation}
 K_\lambda^{\mathrm{mech}}(q)=q^2\mathcal E_\lambda(q),
 \qquad \mathcal E_\lambda(q)\sim|q|\to0,
 \label{eq:scaleDependentModulus}
\end{equation}
where $\mathcal E_L$ and $\mathcal E_T$ reduce to the longitudinal and shear
moduli in the isotropic continuum limit.  The defining mechanical signature
would therefore be a modulus that vanishes linearly at long wavelength.  A
weak external force provides a direct test because the static susceptibility
depends on the response operator, not on the spontaneous noise amplitude:
\begin{equation}
 \avg{u_\lambda(\vq)}=\chi_\lambda(q)f_\lambda^{\mathrm{ext}}(\vq),
 \qquad
 K_{\mathrm{resp},\lambda}(q)=\chi_\lambda^{-1}(q).
 \label{eq:mechanicalResponseTest}
\end{equation}
Probe amplitudes $A =5\times10^{-4}$, $10^{-3}$, and $2\times10^{-3}$ are used to
verify that the measurement remains in the linear-response regime.}

\SK{Figure~\ref{fig:modeT}(a) shows the resulting static response stiffness. The
three probe amplitudes give indistinguishable values of $K_{\mathrm{resp},\lambda}(q)$ within numerical resolution, demonstrating that the response is linear for activity strength $f_0=2.0$. More importantly, both the longitudinal and transverse stiffnesses follow
\begin{equation}
K_{\mathrm{resp},\lambda}(q)\sim q^2
\end{equation}
over the accessible low-$q$ range and are inconsistent with a cubic mechanical response.}
We therefore conclude that the nonanalytic $\Hcov_\lambda(q)\sim|q|^3$ operator characterizes the stationary nonequilibrium covariance but is not the mechanical restoring operator. 
\SK{Accordingly, the additional infrared enhancement of $\Cu_\lambda(q)$ cannot be attributed to nonlocal elastic softening and must instead originate from the spatial and temporal structure of the nonequilibrium fluctuations. This exclusion transfers the unexplained infrared factor from the denominator of Eq.~\eqref{eq:diagnostic-decomposition} to the numerator, motivating the second hypothesis.}

\begin{figure*}[t]
 \centering
  \begin{overpic}[width=0.48\linewidth]{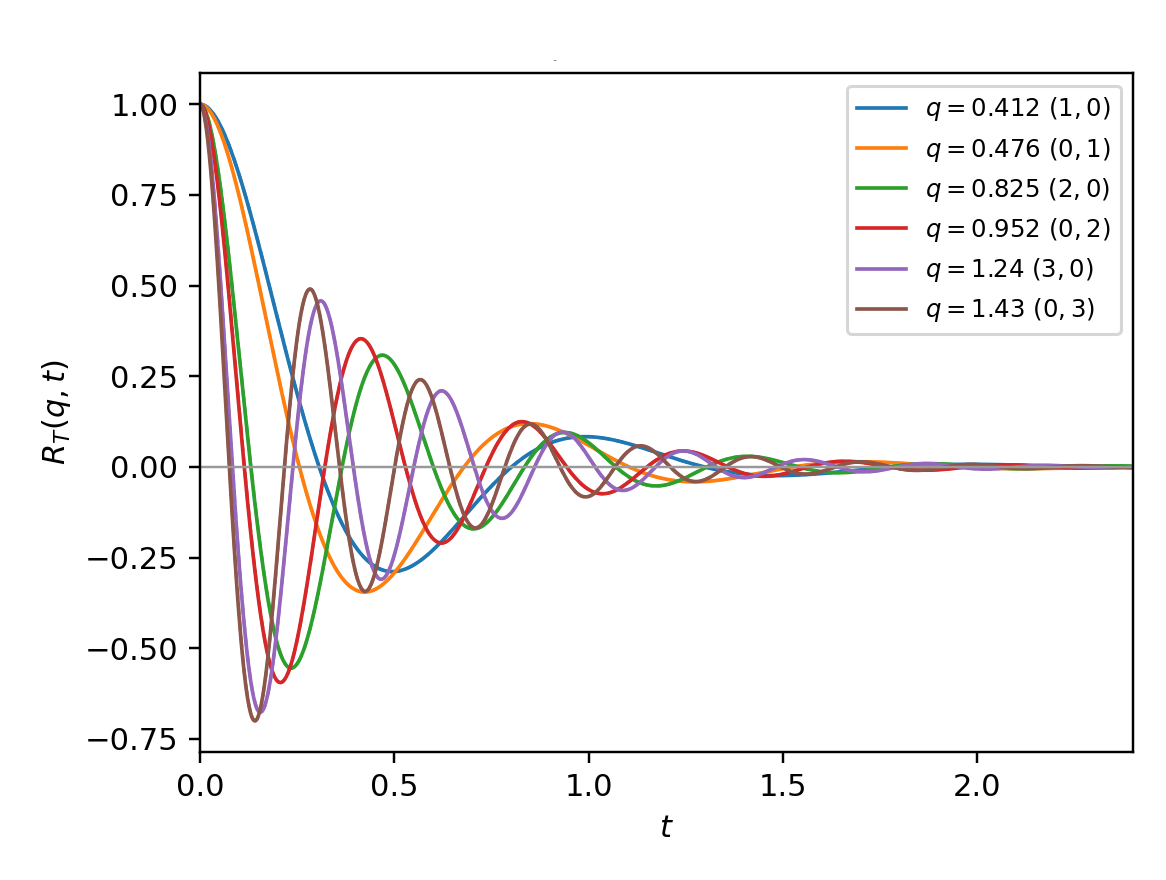}
    \put(80,20){\large\bfseries (a)}
    \put(78,42){\textcolor{red}{$f_0=2.0$}}
\end{overpic}
\begin{overpic}[width=0.48\linewidth]{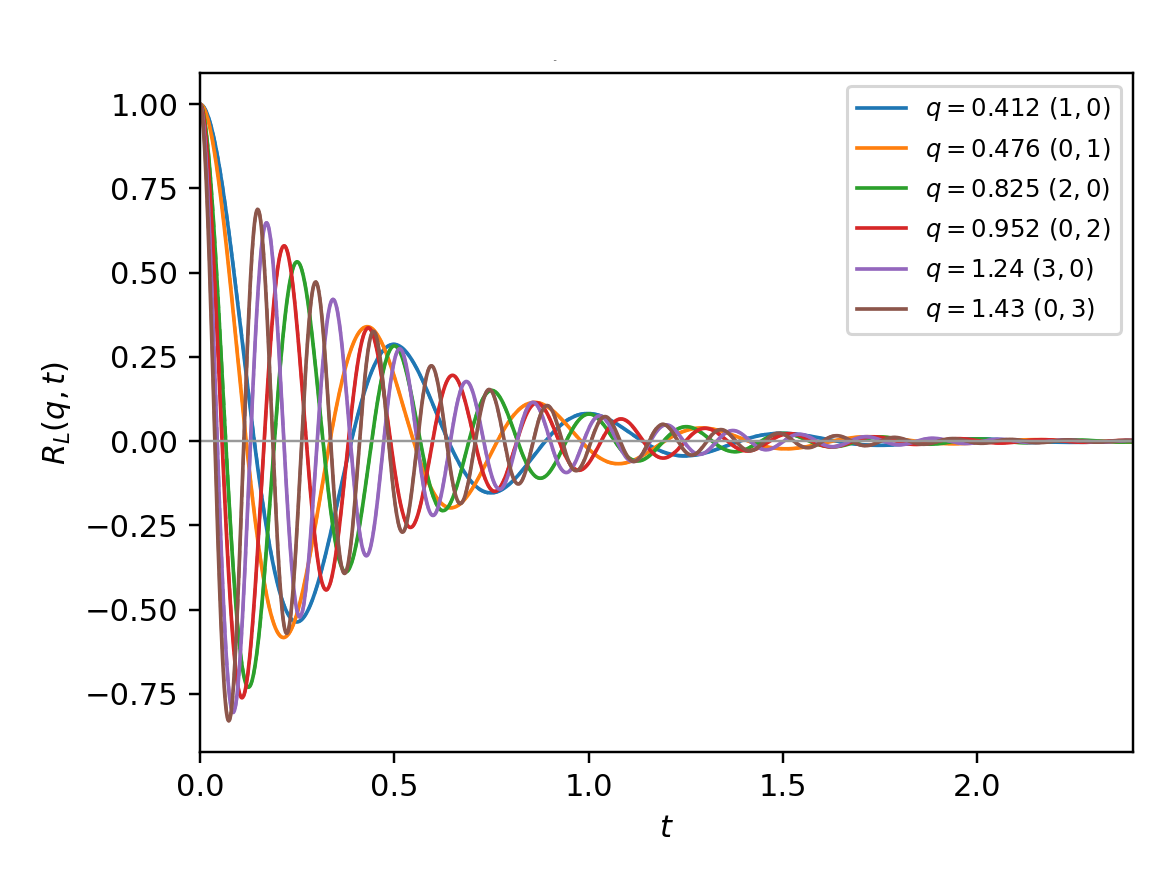}
    \put(80,20){\large\bfseries (b)}
    \put(78,42){\textcolor{red}{$f_0=2.0$}}
\end{overpic}
\begin{overpic}[width=0.48\linewidth]{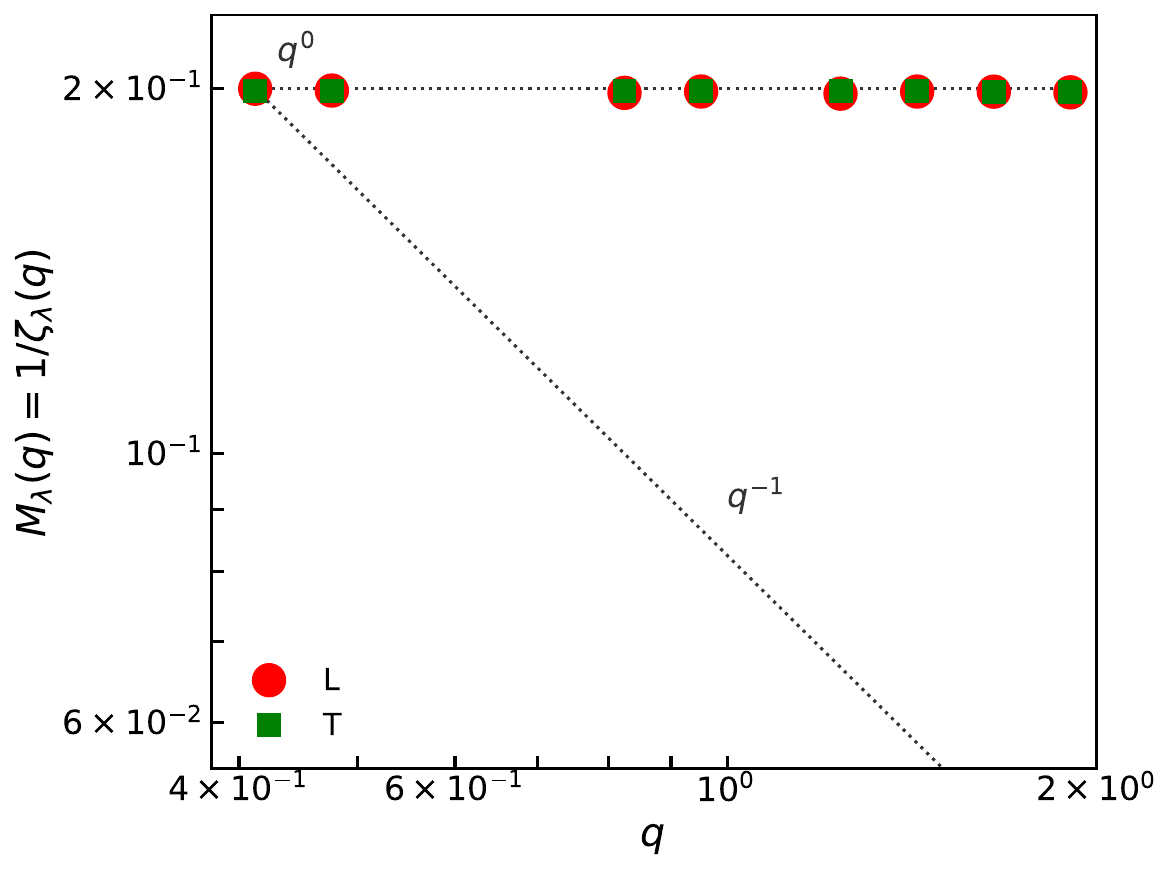}
    \put(21,53){\large\bfseries (c)}
    \put(78,15){\textcolor{red}{$f_0=2.0$}}
\end{overpic}
\begin{overpic}[width=0.48\linewidth]{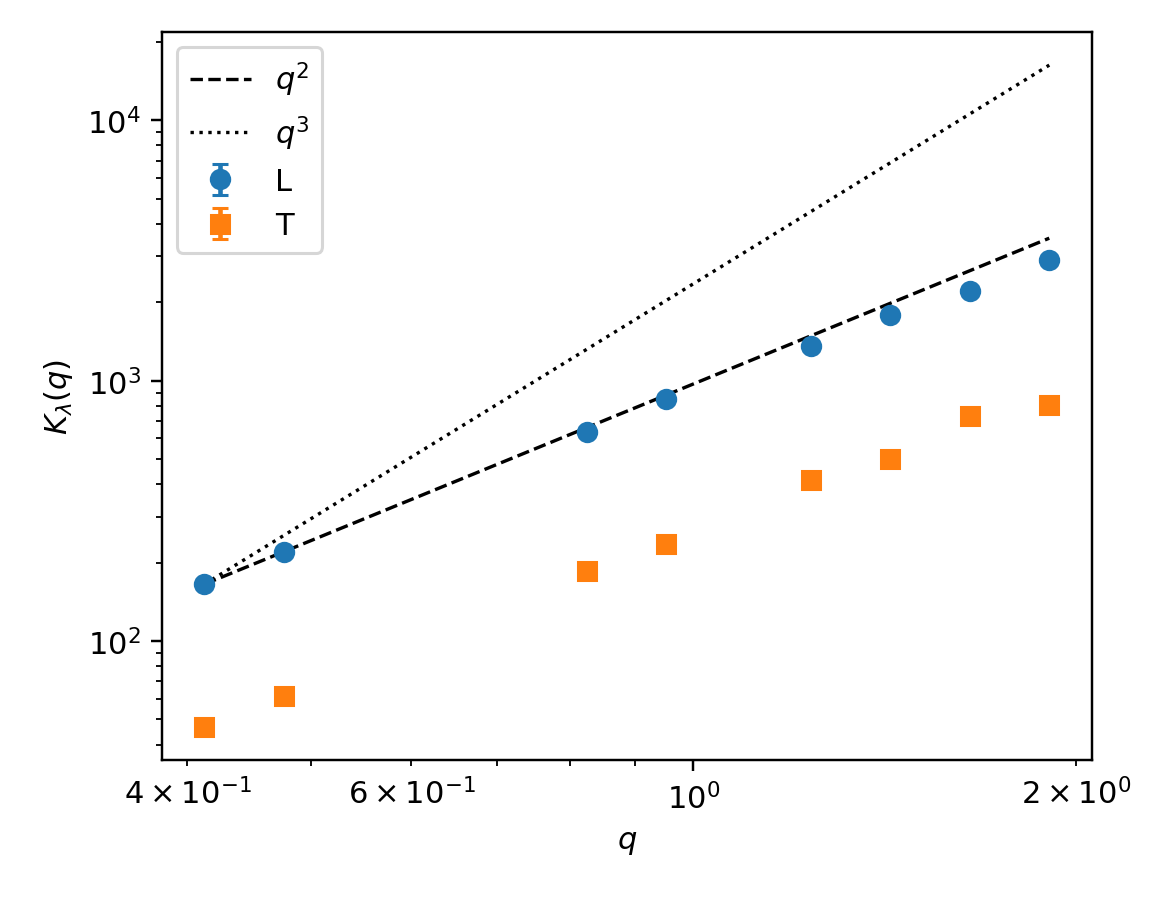}
    \put(30,67){\large\bfseries (d)}
    \put(78,15){\textcolor{red}{$f_0=2.0$}}
\end{overpic}

 \caption{Tests of nonlocal kinetic mechanisms. (a,b) Normalised response $R_\lambda(q,t)$ for different low-$q$ transverse and longitudinal modes, respectively, at activity strength $f_0=2.0$. (c) Wave-vector-dependent mobility $\mathcal M_\lambda(q)=1/\zeta_\lambda(q)$ obtained from the dynamic response. Both longitudinal and transverse mobilities remain approximately constant,
consistent with the $q^0$ guide and inconsistent with the singular $\mathcal M_\lambda\sim q^{-1}$ behavior for $f_0=2.0$. (d) Dynamic stiffness $K_\lambda$ fitted from the response function.}

 \label{fig:failed}
\end{figure*}

\subsection{Scale-dependent active mode temperature}
\label{sec:mode-temperature}

\SK{The effective mode-dependent temperature can be defined, using the independently measured response kernel, as
\begin{equation}
 T_{\mathrm{mode},\lambda}(q)
 \equiv K_{\mathrm{resp},\lambda}(q)\Cu_\lambda(q).
 \label{eq:Tmode}
\end{equation}
This is the same diagnostic quantity denoted by
$T_{\mathrm{eff},\lambda}$ in
Eq.~\eqref{eq:diagnostic-decomposition}.  Using
$K_{\mathrm{resp},\lambda}\simeq M_\lambda q^2$ and
$\Cu_\lambda\simeq A_\lambda q^{-3}$ gives
\begin{equation}
  T_{\mathrm{mode},\lambda}(q) \simeq M_\lambda A_\lambda q^{-1}.
  \label{eq:modeTemperatureAmplitude}
\end{equation}
$T_{\mathrm{mode},\lambda}$ is an energy-valued fluctuation-response ratio, equal to $k_{\mathrm B}T$ at equilibrium. Thus, the long-wavelength modes fluctuate as if their effective energy scale increases with wavelength.  This does not define a thermodynamic temperature
\cite{Loi2008,Cugliandolo2011,Szamel2014}: different polarizations may have different amplitudes, and no zeroth-law or fluctuation--dissipation property follows from Eq.~\eqref{eq:Tmode}.}

\SK{Figure~\ref{fig:modeT}(b) shows $T_{\mathrm{mode},\lambda}(q)$ obtained by combining the measured response kernel with the displacement covariance for $f_0=2.0$. Both polarizations exhibit an
enhancement toward small $q$, with a substantially larger amplitude in the
transverse sector. The transverse data are approximately compatible with a
$q^{-1}$ dependence over the low $q$ regime. The longitudinal data show a weaker increase, and therefore do not establish a clean inverse-$q$ scaling. At larger $q$, both polarizations approach the thermal background scale, $T_{\mathrm{mode},\lambda}\simeq k_{\rm B}T$, indicating that the
nonequilibrium enhancement is concentrated predominantly in the
long-wavelength modes.}

\SK{It is important that Eq.~\eqref{eq} is constructed from the same
response and covariance whose scaling we seek to understand. Consequently,
the increase of $T_{\mathrm{mode},\lambda}(q)$ is not an independent
microscopic explanation of $\Cu_\lambda(q)\sim q^{-3}$; it is an alternative representation of the anomalous fluctuation--response ratio. The result shows that activity produces a scale- and polarization-dependent violation of the equilibrium fluctuation--dissipation relation
\cite{Fodor2016,DalCengio2019,Mandal2017}, but it does not identify
the physical process generating that violation. We therefore next examine
whether the enhancement originates from nonlocal mobility, spatially correlated forcing, or collective interaction-stress fluctuations.}

\subsection{Nonlocal mobility}
\label{sec:nonlocal-mobility}
\SK{The preceding static-response analysis finds an elastic stiffness $K_{\mathrm{resp},\lambda}(q)\sim q^2$ over the measured wave-vector range. A distinct possible source of the anomalous covariance is a nonlocal kinetic coefficient. A spatially nonlocal friction law that is local
in time can be written schematically as
\begin{equation}
\begin{split}
 m\ddot u_\alpha(\bm r,t)
 +&\int \mathrm d\bm r'\,
 \zeta_{\alpha\beta}(\bm r-\bm r')\dot u_\beta(\bm r',t) \\
 &=F_\alpha^{\mathrm{int}}(\bm r,t)
 +f_\alpha^a(\bm r,t)+\xi_\alpha(\bm r,t).
 \label{eq:nlm-real-space}
\end{split}
\end{equation}
Here $\alpha,\beta$ denote Cartesian components, with summation over $\beta$, and $\bm F^{\mathrm{int}}$ is conservative force, $\bm f^a$ is the active force, and $\bm\xi$ is the thermal random force. For local drag, $\zeta_{\alpha\beta}(\bm r-\bm r')=
\gamma\delta_{\alpha\beta}\delta(\bm r-\bm r')$, a spatially extended friction couples velocities at different positions \cite{baumgartl2008phonon}.}

\SK{Assuming longitudinal and transverse modes decouple, spatial Fourier transformation gives the damping $\zeta_\lambda(q)$, where $q=|\bm q|$ and $\lambda=L,T$. We define its inverse as the effective mobility,
\begin{equation}
 \mathcal M_\lambda(q)=\frac{1}{\zeta_\lambda(q)}.
 \label{eq:nlm-mobility}
\end{equation}
For local drag, $\mathcal M_\lambda=1/\gamma$. With
$F_\lambda^{\mathrm{el}}(\bm q,t)= -K_{\mathrm{resp},\lambda}(q)u_\lambda(\bm q,t)$,
the linear mode equation becomes
\begin{equation}
\begin{split}
 m\ddot u_\lambda(\bm q,t)
 +\zeta_\lambda(q)\dot u_\lambda(\bm q,t)
 +&K_{\mathrm{resp},\lambda}(q)u_\lambda(\bm q,t) \\
 =&f_\lambda^a(\bm q,t)+\xi_\lambda(\bm q,t).
 \label{eq:nlm-mode-dynamics}
\end{split}
\end{equation}
Now, using the temporal Fourier transform, one can write
$u_\lambda(\bm q,\omega)=\int\mathrm dt\,e^{i\omega t}u_\lambda(\bm q,t)$,
then the displacement susceptibility is
\begin{equation}
 \chi_\lambda(q,\omega)=
 \frac{1}{K_{\mathrm{resp},\lambda}(q)-m\omega^2
 -i\omega\zeta_\lambda(q)}.
 \label{eq_sus}
\end{equation}}

\SK{To illustrate how damping could affect the equal-time fluctuations,
consider an effective white-noise approximation for the active force,
\begin{equation}
 \left\langle f_\lambda^a(\bm q,t)
 [f_\lambda^a(\bm q,t')]^*\right\rangle
 =2D_\lambda^f(q)\delta(t-t'),
 \label{eq:nlm-white-active-force}
\end{equation}
where $D_\lambda^f(q)$ is the active force-noise intensity. Defining the power spectrum as the temporal Fourier transform of the stationary correlation function, the active contribution to the displacement spectrum is
\begin{equation}
\begin{split}
 S_\lambda^{u,a}(q,\omega)
 &=2D_\lambda^f(q)|\chi_\lambda(q,\omega)|^2 \\
 &=\frac{2D_\lambda^f(q)}{
 [K_{\mathrm{resp},\lambda}(q)-m\omega^2]^2
 +\omega^2\zeta_\lambda^2(q)}.
 \label{eq:nlm-white-spectrum}
\end{split}
\end{equation}
For a stable nonzero mode with positive stiffness and damping, the
corresponding equal-time covariance is
\begin{align}
 C_\lambda^{u,a}(q)
 &=\int_{-\infty}^{\infty}\frac{\mathrm d\omega}{2\pi}
 S_\lambda^{u,a}(q,\omega)\nonumber\\
 &=\frac{D_\lambda^f(q)}{
 K_{\mathrm{resp},\lambda}(q)\zeta_\lambda(q)}
 =\frac{D_\lambda^f(q)\mathcal M_\lambda(q)}{
 K_{\mathrm{resp},\lambda}(q)}.
 \label{Cu_mobility_relation}
\end{align}}

\SK{Within the white-noise approximation, if
$K_{\mathrm{resp},\lambda}(q)\sim q^2$ and
$D_\lambda^f(q)\sim q^0$, an active covariance
$C_\lambda^{u}(q)\sim q^{-3}$ would require
\begin{equation}
 \mathcal M_\lambda(q)\sim q^{-1},
 \qquad\text{equivalently}\qquad
 \zeta_\lambda(q)\sim q.
 \label{eq:nlm-required-mobility}
\end{equation}
In this scenario, long-wavelength modes would experience progressively
weaker damping.}

\SK{We measure the mobility using paired free-relaxation trajectories. Starting from the same saved configuration, the perturbed
replica receives the initial displacement
\begin{equation}
 \Delta\bm r_i(0)
 =A\hat{\bm e}_\lambda
 \cos(\bm q\cdot\bm R_i),
 \qquad\Delta\dot{\bm r}_i(0)=0,
 \label{eq:nlm-initial-perturbation}
\end{equation}
where $i$ labels particles, $A$ is a displacement
amplitude, $\bm R_i$ is the fixed reference lattice position, and
$\hat{\bm e}_\lambda$ is the longitudinal or transverse unit vector.
No additional probe force is applied during relaxation. Both replicas
retain the original activity and thermal noise, with identical
orientation histories and thermal kicks; after applying the probe force, we let them relax. The explicit active and
thermal forces therefore cancel in their difference, while the
interaction response remains dependent on the active background.}

\SK{Let $\Delta\bm r_i^{\mathrm u}(t)$ denote the difference of unwrapped
positions between the perturbed and reference replicas, and define
$\Delta\bm r_{\mathrm{CM}}(t)=
N^{-1}\sum_j\Delta\bm r_j^{\mathrm u}(t)$, where $N$ is the number
of particles. We measure the complex amplitude
\begin{equation}
 a_\lambda(\bm q,t)=\frac{1}{N}\sum_i
 \hat{\bm e}_\lambda\cdot
 [\Delta\bm r_i^{\mathrm u}(t)-\Delta\bm r_{\mathrm{CM}}(t)]
 e^{-i\bm q\cdot\bm R_i}
 \label{eq:nlm-difference-mode}
\end{equation}
and its signed normalized response
\begin{equation}
 R_\lambda(q,t)=
 \frac{\operatorname{Re}[a_\lambda(\bm q,t)
 a_\lambda^*(\bm q,0)]}{|a_\lambda(\bm q,0)|^2}.
 \label{eq:nlm-normalized-response}
\end{equation}
Any constant Fourier normalization cancels in this ratio.
We average the response over the sampled initial states and fit it
to the effective single-mode equation
\begin{equation}
 m\ddot R_\lambda(q,t)
 +\zeta_\lambda(q)\dot R_\lambda(q,t)
 +K_{\mathrm{dyn},\lambda}(q)R_\lambda(q,t)=0,
 \label{eq:nlm-relaxation-fit}
\end{equation}
with $R_\lambda(q,0)=1$ and $\dot R_\lambda(q,0)=0$.
The mass is fixed at its simulation value, while the positive damping
$\zeta_\lambda$ and dynamic stiffness $K_{\mathrm{dyn},\lambda}$
are fitted. In particular, the fitted damping is not fixed to the
microscopic drag coefficient $\gamma$, and the fitted stiffness is
not fixed to the independently measured $K_{\mathrm{resp},\lambda}$.
The fitting procedure includes underdamped, critically damped, and
overdamped solutions. In the underdamped regime,
\begin{align}
 R_\lambda(q,t)&=e^{-\alpha_\lambda t}
 \left[\cos(\omega_{d,\lambda}t)
 +\frac{\alpha_\lambda}{\omega_{d,\lambda}}
 \sin(\omega_{d,\lambda}t)\right],\nonumber\\
 \alpha_\lambda&=\frac{\zeta_\lambda}{2m},\qquad
 \omega_{d,\lambda}^2=
 \frac{K_{\mathrm{dyn},\lambda}}{m}-\alpha_\lambda^2.
 \label{eq:nlm-underdamped-response}
\end{align}
Here $\alpha_\lambda$ is the decay rate of the envelope and
$\omega_{d,\lambda}$ is the damped oscillation frequency.
The mobility is obtained as $\mathcal M_\lambda(q)=1/\zeta_\lambda(q)$.}

\begin{figure*}[t]
 \centering
 \begin{overpic}[width=0.48\linewidth]{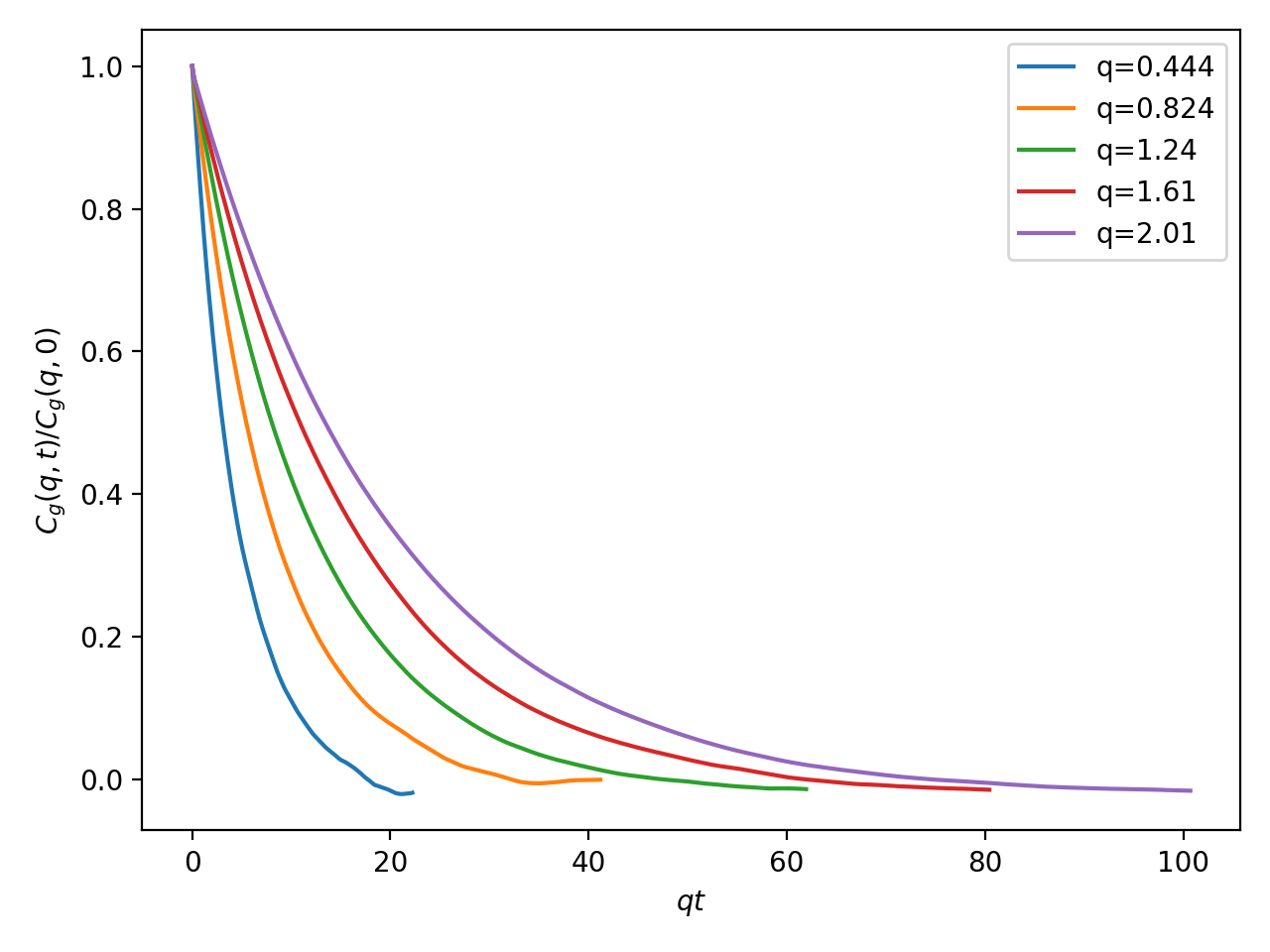}
    \put(27,63){\large\bfseries (a)}
    \put(43,67){Transverse}
    \put(81,50){\textcolor{red}{$f_0=2.0$}}
\end{overpic}
\begin{overpic}[width=0.48\linewidth]{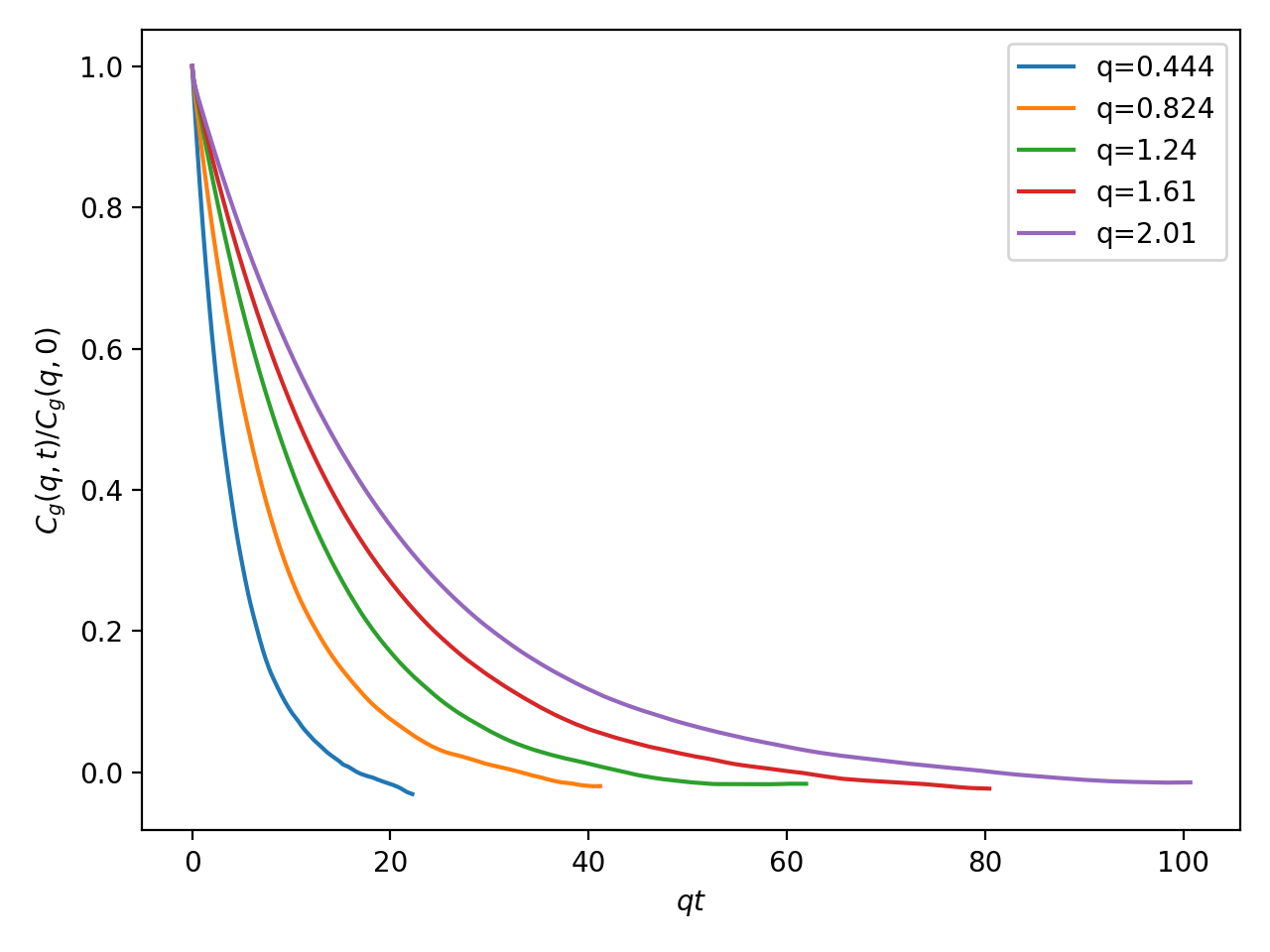}
    \put(25,63){\large\bfseries (b)}
    \put(43,67){Longitudinal}
    \put(81,50){\textcolor{red}{$f_0=2.0$}}
\end{overpic}
\begin{overpic}[width=0.48\linewidth]{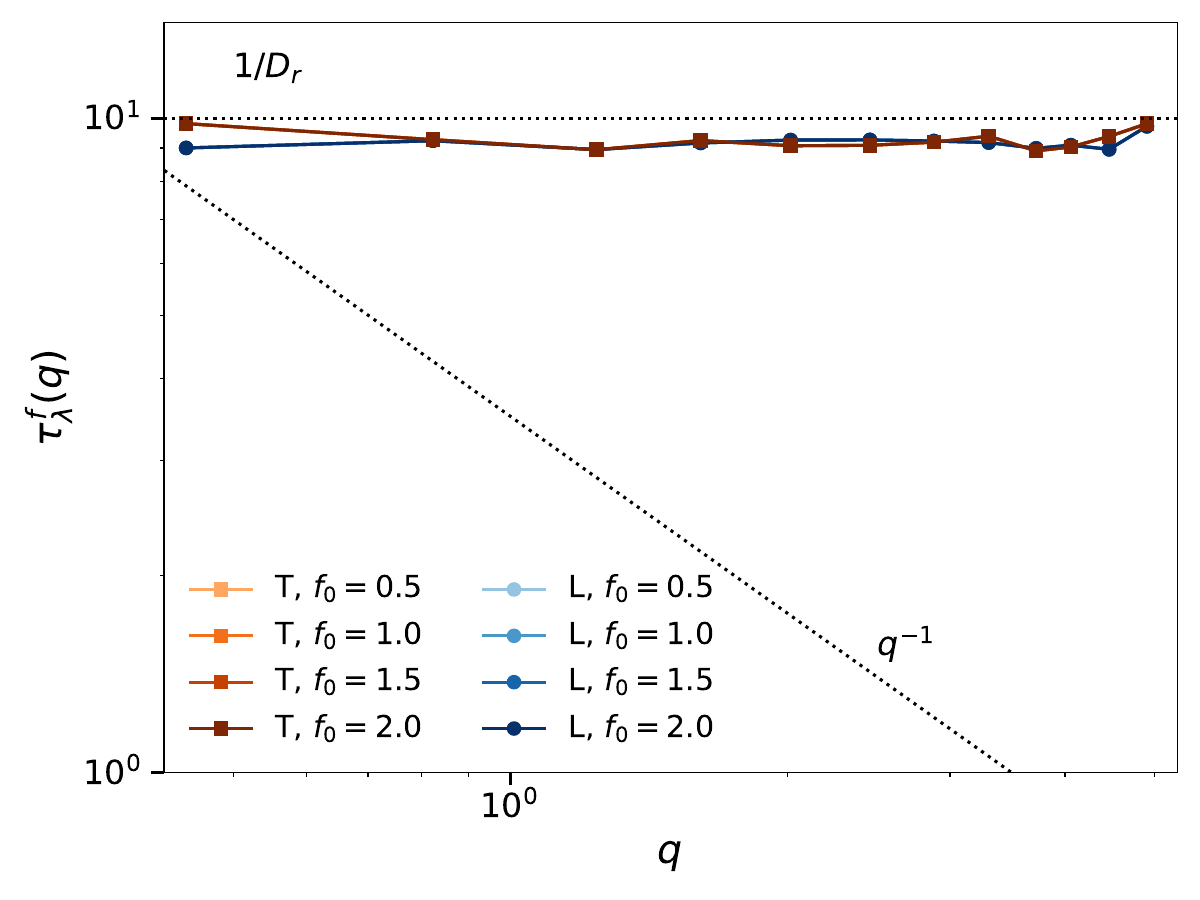}
    \put(84,15){\large\bfseries (c)}
    \put(55, 27){\includegraphics[width=0.2\textwidth]{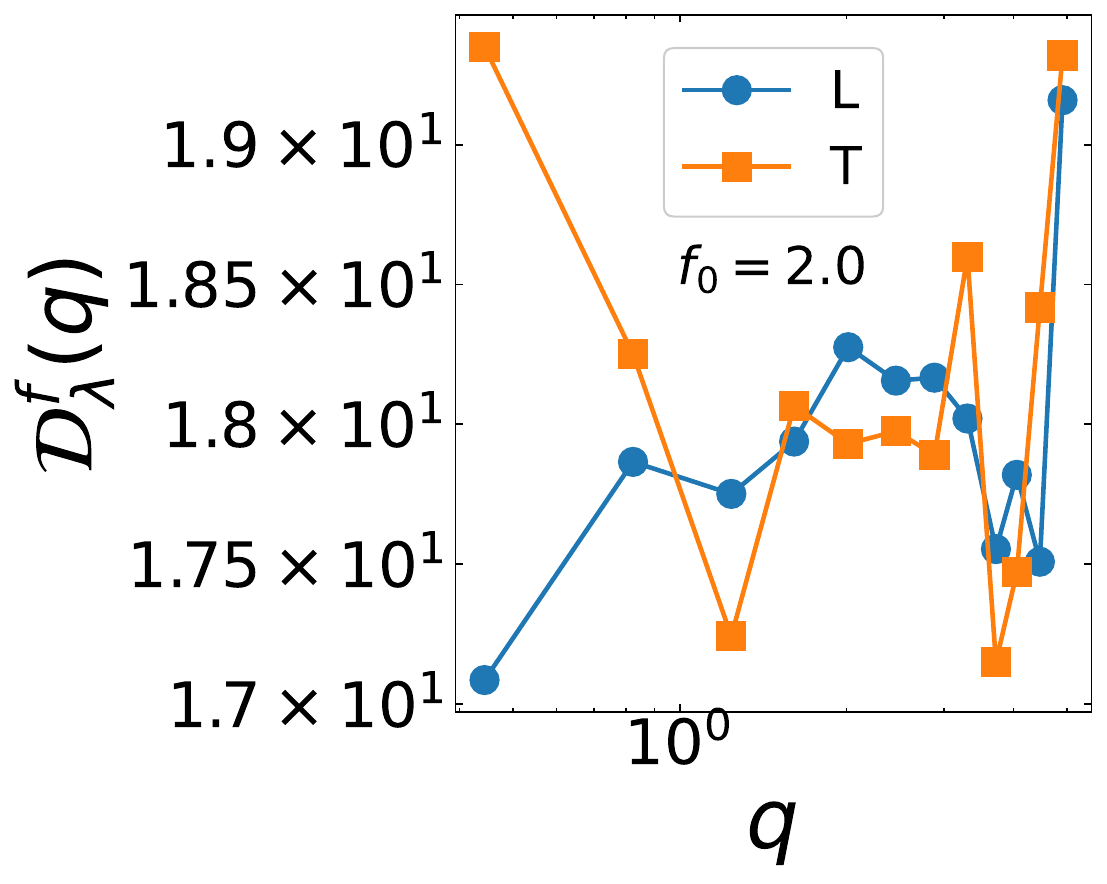}}
\end{overpic}
\begin{overpic}[width=0.48\linewidth]{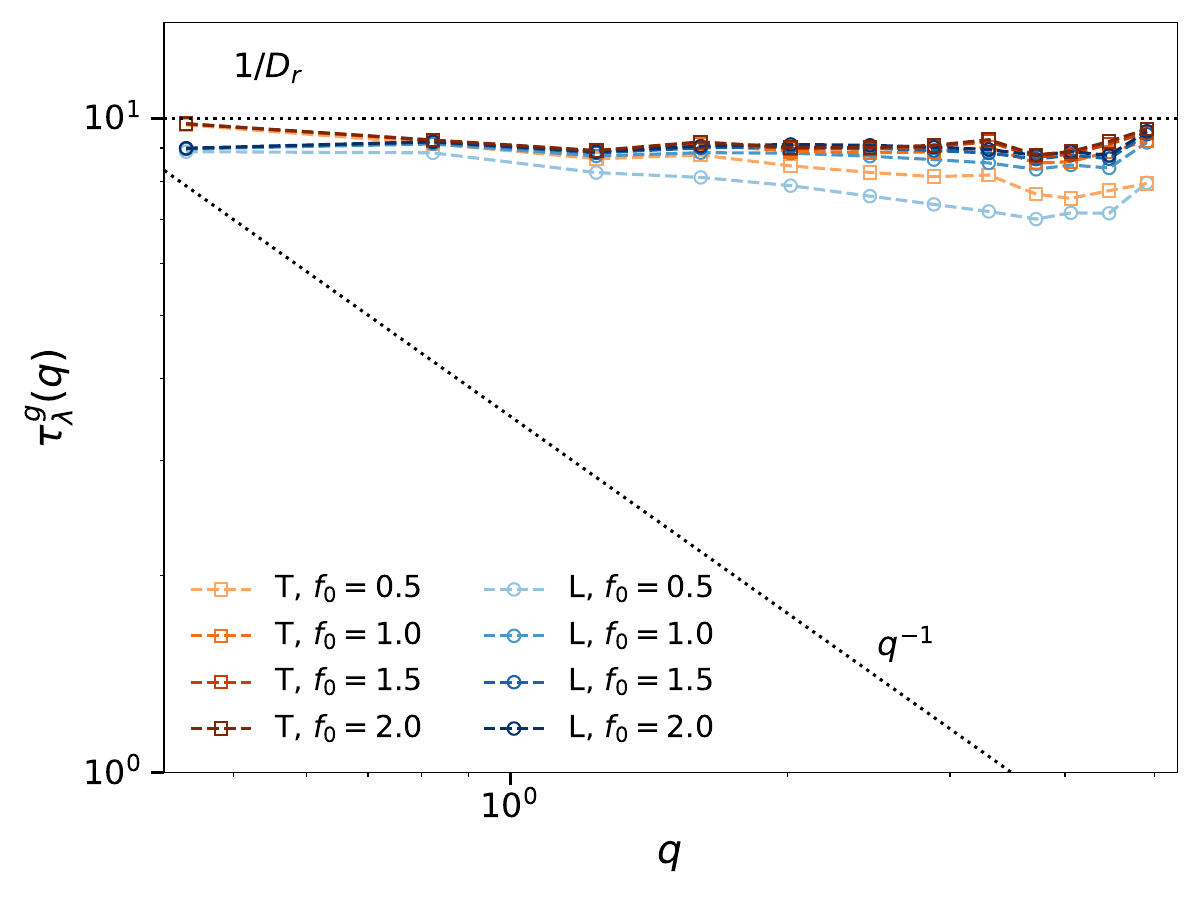}
    \put(85,15){\large\bfseries (d)}
    \put(55, 22){\includegraphics[width=0.2\textwidth]{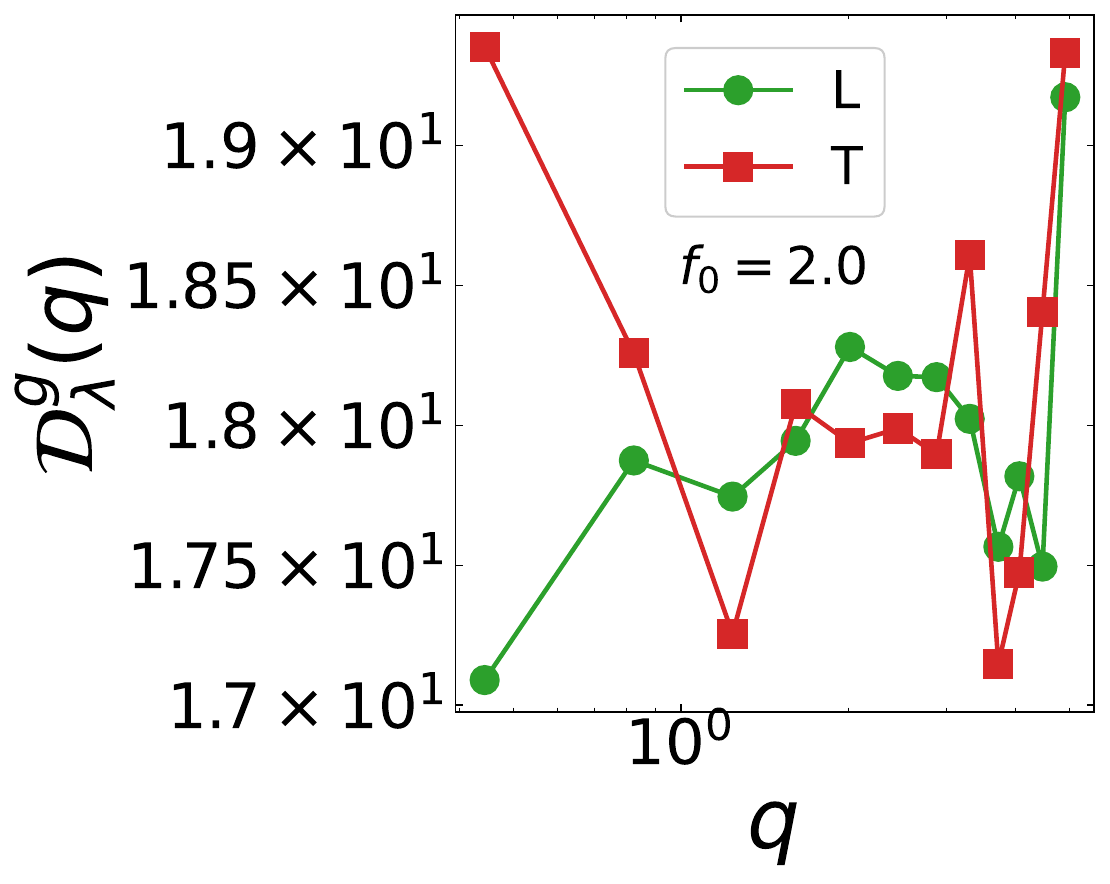}}
\end{overpic}
\caption{\label{fig:force}
Normalised $g_\lambda$ correlation for (a) transverse and (b) longitudinal modes as a function of time for different low-$q$ values at activity strength $f_0=2.0$.  
(c) Correlation time $\tau_\lambda^f(q)$ of the bare active-force modes for different propulsion strengths. The longitudinal and transverse results remain close to the single-particle persistence time $D_r^{-1}$ and do not follow the $q^{-1}$ guide. (c) Correlation time $\tau_\lambda^g(q)$ of the effective residual force $g_\lambda=f_\lambda^a+(f_\lambda^{\mathrm{int}}-f_\lambda^{\mathrm{harm}})$. Insets: integrated autocorrelation intensities, $D_\lambda(q)$, for active and residual force with activity strength $f_0=2.0$ for longitudinal and transverse modes.}

\end{figure*}


\SK{Figure~\ref{fig:failed}(a) and (b) show the normalised response for different low-$q$ modes as a function of time at activity strength $f_0=2.0$. Then Figure~\ref{fig:failed}(c) shows mobility extracted from these free-relaxation fits. Over the measured wave-vector range and within the fitted time interval, the longitudinal and transverse results
are approximately wave-vector independent and remain close to
$\mathcal M_\lambda(q)\simeq\gamma^{-1}=0.2$ for $\gamma=5$.
They therefore support $\mathcal M_\lambda(q)\sim q^0$ and show no
evidence for the $q^{-1}$ enhancement in
Eq.~\eqref{eq:nlm-required-mobility}. Dynamical response fit in Figure~\ref{fig:failed}(d) also supports the same argument, and we see $q^2$ divergence.}

\SK{This behaviour is consistent with the local momentum dissipation and
absence of solvent-mediated hydrodynamic interactions in the dry
active-particle model \cite{Ramaswamy2010,Marchetti2013,Bechinger2016}.
Within the wave-vector range and time interval tested by this
effective damping model, the data do not support a singular mobility
as the origin of the additional infrared enhancement of the covariance.
Together with the elastic static stiffness, this motivates examining
the spatial and temporal correlations of the nonequilibrium force
and stress fluctuations.}

\begin{figure*}[t]
 \centering
 \begin{overpic}[width=0.48\linewidth]{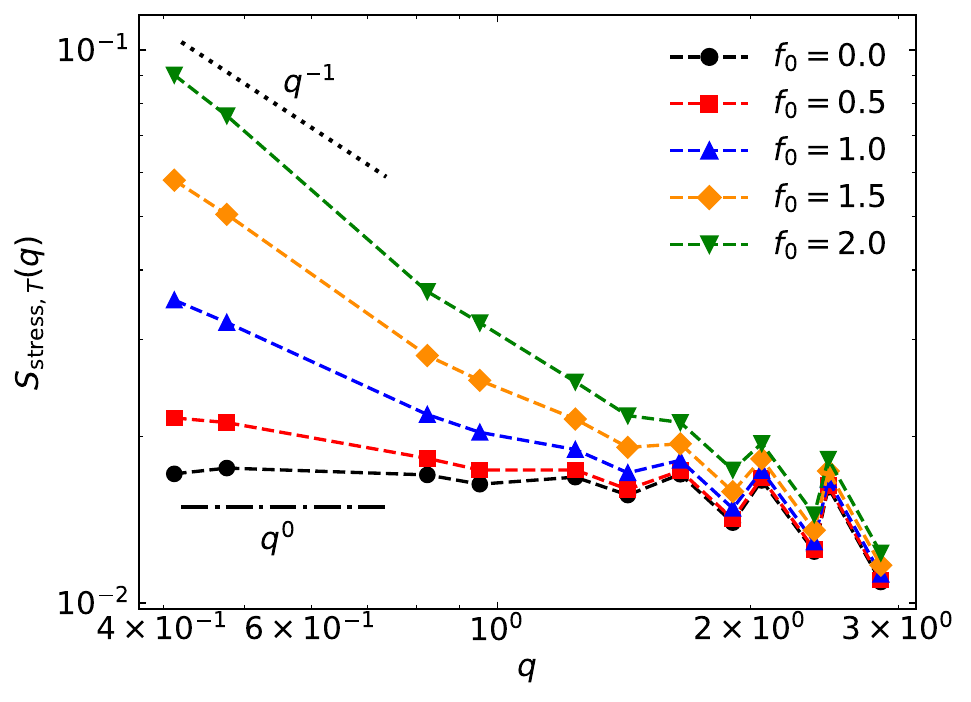}
    \put(45,63){\large\bfseries (a)}
\end{overpic}
\begin{overpic}[width=0.48\linewidth]{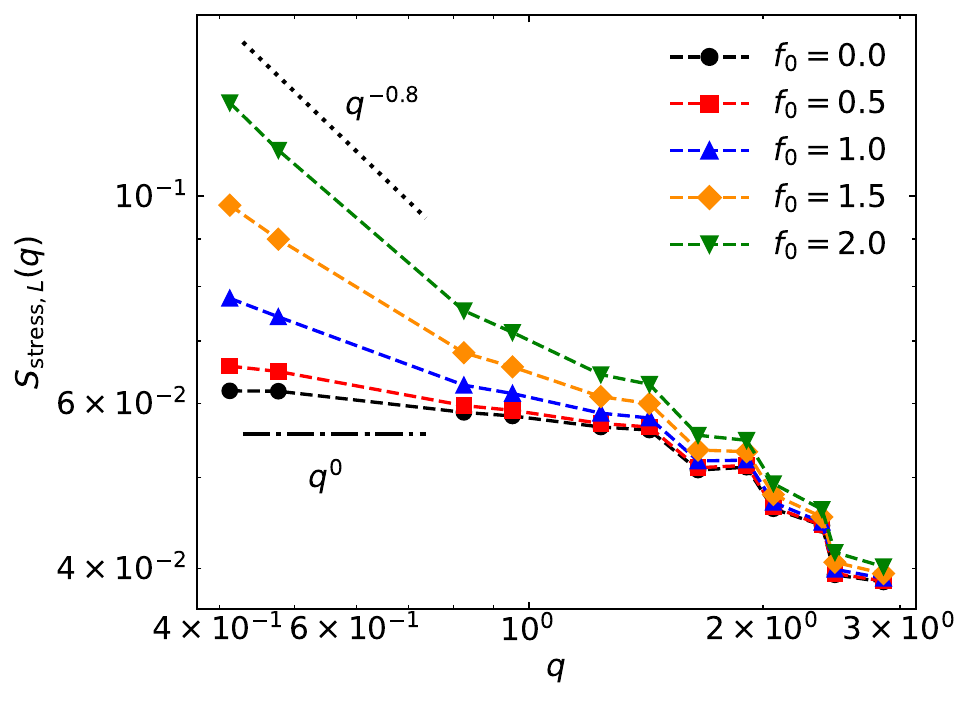}
    \put(47,63){\large\bfseries (b)}
\end{overpic}
\begin{overpic}[width=0.48\linewidth]{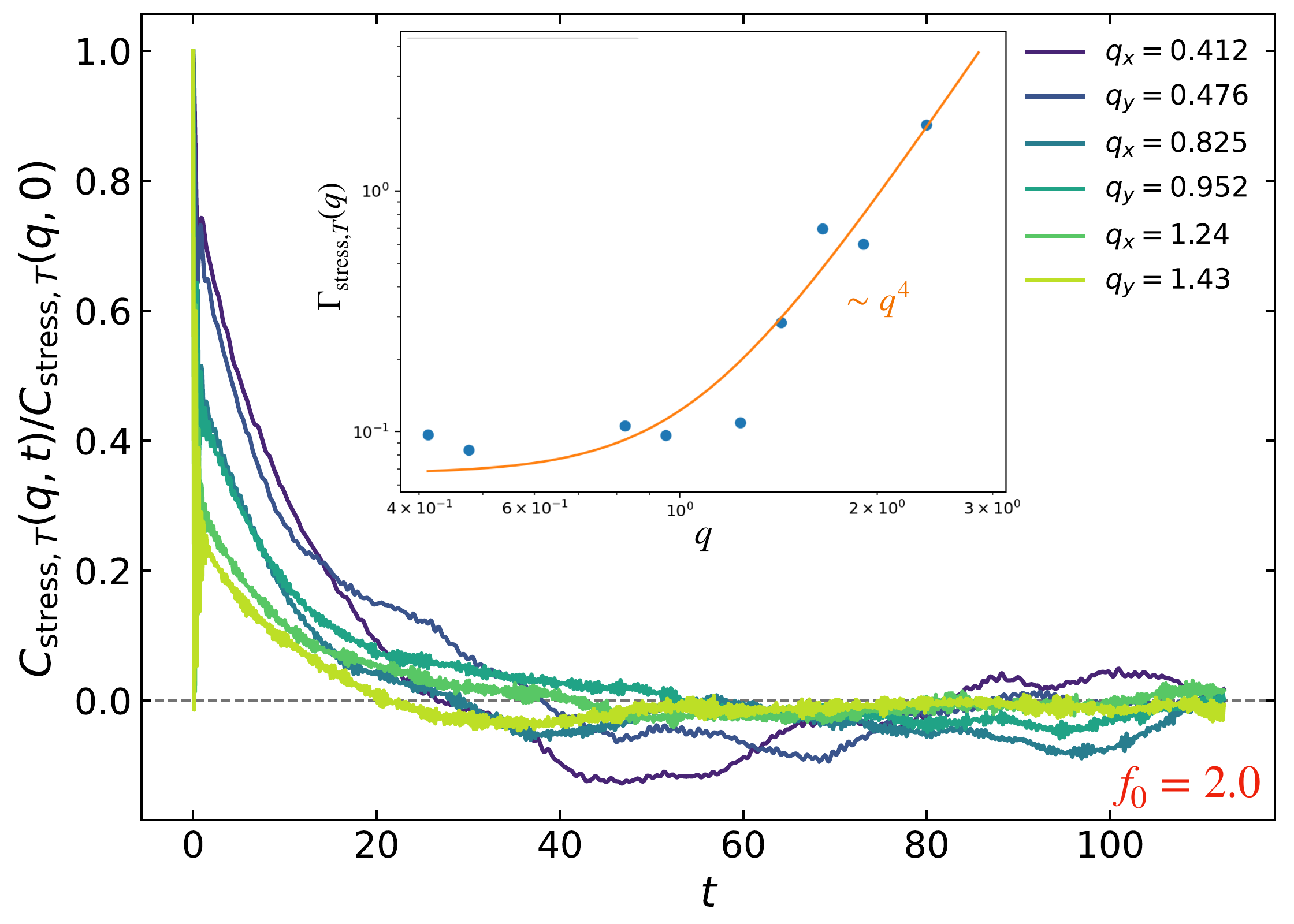}
    \put(13,11){\large\bfseries (c)}
\end{overpic}
\begin{overpic}[width=0.48\linewidth]{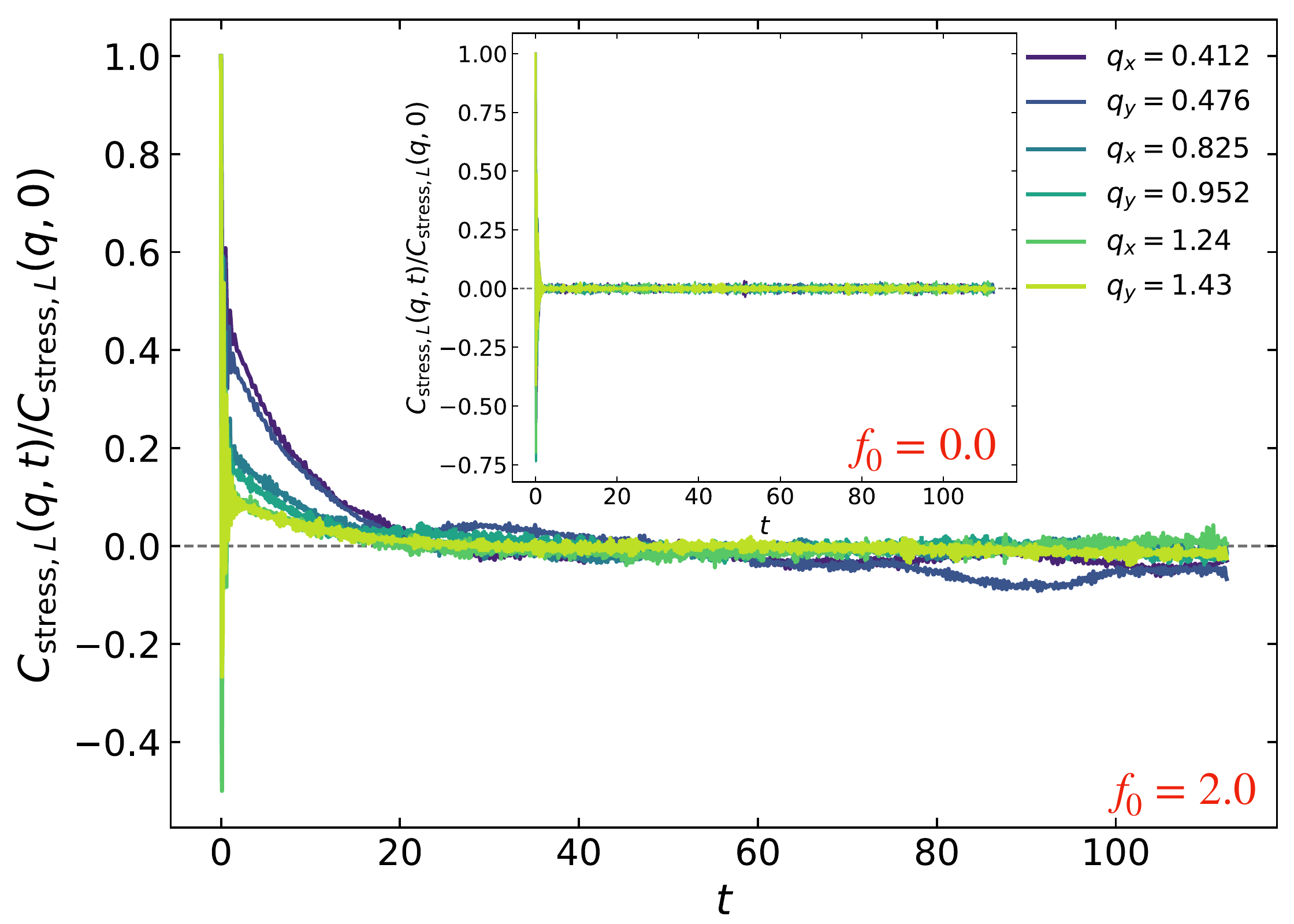}
    \put(19,11){\large\bfseries (d)}
\end{overpic}
\caption{\label{fig:stress}
Interaction-stress fluctuations and their temporal correlations.
(a,b) Equal-time spectra of the transverse shear component,
$S_{\mathrm {stress},T}(q)$, and the longitudinal normal component,
$S_{\mathrm {stress},L}(q)$, respectively, measured from the conservative
pair forces for different propulsion strengths $f_0$. In the passive system, both spectra are approximately independent of $q$, whereas
activity enhances the long-wavelength fluctuations. At $f_0=2.0$, the
transverse spectrum follows approximately $q^{-1}$ and the longitudinal
spectrum approximately $q^{-0.8}$ over the low-$q$ range.
The dotted and dash-dotted lines are the indicated power-law guides.
(c) Normalized connected autocorrelation functions
$C_{\mathrm{stress},T}(q,t)/C_{\mathrm{stress},T}(q,0)$ for the transverse interaction-stress component. The main panels show the active system at $f_0=2.0$. Inset show relaxation rate $\Gamma_{\mathrm{stress},T}(q)$ as a function of $q$, where $C_{\mathrm{stress},L}(q,t)/C_{\mathrm{stress},L}(q,0) \propto e^{-\Gamma(q)t}$. (d) Normalized connected autocorrelation functions
$C_{\mathrm{stress},L}(q,t)/C_{\mathrm{stress},L}(q,0)$ for the longitudinal interaction-stress component. The main panels show the active system at $f_0=2.0$, while the
insets show the passive system. Colors denote wave vectors directed
along the $x$ and $y$ axes, as specified in the legends. Passive stress
correlations decay rapidly on a microscopic time scale, whereas activity
produces a pronounced slowing of the long-wavelength stress modes, with the
smallest-$q$ modes retaining correlations for the longest times.
}
\end{figure*}

\subsection{Wave-vector-dependent active-force correlations}
\label{sec:active-force-correlations}

\SK{For a general colored force, define its equal-time spectrum, normalized
correlation function, and correlation time by Eq.~\eqref{eq:Cf},\eqref{eq:tauf} and \eqref{eq:Df} respectively. Now, using Eq.~\eqref{Cu_mobility_relation} with $K\sim q^2$ and $\mathcal M\sim q^0$, the observed $q^{-3}$ spectrum would require
\begin{equation}
 \mathcal D_\lambda^f(q)\sim q^{-1}.
 \label{eq:requiredForceIntensity}
\end{equation}
If $S_\lambda^f\sim q^{-a}$ and
$\tau_\lambda^f\sim q^{-b}$, the necessary condition is $a+b=1$.
This separates a spatial-amplitude mechanism ($a>0$) from a collective
persistence mechanism ($b>0$).}

\SK{For the torque-free ABP propulsion, neither divergence is expected
\cite{Bechinger2016,Fily2014}.  In
$d$ dimensions the independent orientation dynamics gives
\begin{equation}
 \avg{n_{i\alpha}(t)n_{j\beta}(0)}
 =\frac{\delta_{ij}\delta_{\alpha\beta}}{d}
 e^{-(d-1)D_r|t|}.
 \label{eq:orientationCorrelationD}
\end{equation}
Using the reference-position Fourier mode defined in Sec.~II B therefore gives
exactly
\begin{align}
 C_\lambda^{f}(q,t)
 &\propto \frac{f_0^2}{d}e^{-(d-1)D_r|t|},\nonumber\\
 S_\lambda^{f}(q)&\sim\frac{f_0^2}{d},\qquad
 \tau_\lambda^{f}=\frac{1}{(d-1)D_r},\nonumber\\
 \mathcal D_\lambda^{f}(q)
 &\sim\frac{f_0^2}{d(d-1)D_r},
 \label{eq:bareActiveForceCorrelation}
\end{align}
independent of $q$.  Thus, uncorrelated propulsion cannot supply the required factor.  Interactions can nevertheless generate a collective effective noise even when the microscopic propulsion is regular. Figure~\ref{fig:force}(c) shows the correlation time of the bare active-force longitudinal and transverse modes for different activity strengths ranging from $0.5$ to $2.0$. For all propulsion strengths and both polarizations,
$\tau_\lambda^f(q)$ remains close to $D_r^{-1}$ and displays no systematic
growth toward small $q$. The integrated intensities reported in Figs.~\ref{fig:force}(c) inset likewise show no $\mathcal D_\lambda^f(q)\sim q^{-1}$ divergence. The results therefore agree with the exact ABP result in Eq.~\eqref{eq:bareActiveForceCorrelation} and are inconsistent with the $q^{-1}$ collective-persistence guide.}

\SK{We next reconstruct the residual force using
\begin{equation}
\begin{split}
 g_\lambda(q,t)
 =&m\ddot u_\lambda+\gamma\dot u_\lambda
 +K_{\mathrm{resp},\lambda}(q)u_\lambda-\xi_\lambda \\
 =&f_\lambda^a(q,t)+f_\lambda^{\mathrm{int}}(q,t)-f_\lambda^{\mathrm{harm}}(q,t)
 \label{eq:effectiveResidual}
\end{split}
\end{equation}
where, $f_\lambda^{\mathrm{harm}}(q,t) = -K_{\mathrm{resp},\lambda}(q)u_\lambda(q,t)$, and $f_\lambda^{\mathrm{int}}$ is conservative interaction force.
It contains the propulsion contribution and
the part of the conservative interaction force not captured by the
linear response kernel. Therefore, it is a reconstructed residual force,
not an independent additive noise. Figure~\ref{fig:force}(a,b) shows the normalised response function of the residual force correlation following Eq.~\ref{eq:Cf} for activity strength $f_0=2.0$ for transverse and longitudinal modes. The corresponding residual-force correlation times are shown in Fig.~\ref{fig:force}(d) for different active strengths for both longitudinal and transverse modes. Interactions produce modest deviations
from the bare active-force values, but $\tau_\lambda^g(q)$ remains finite and
only weakly dependent on wave number. In particular, neither polarization
approaches $\tau_\lambda^g\sim q^{-1}$. The integrated intensities reported in Figs.~\ref{fig:force}(d) inset likewise show no $\mathcal D_\lambda^g(q)\sim q^{-1}$ divergence.}

\SK{These measurements rule out a genuine infrared divergence arising from an
independently prescribed additive force with either a singular equal-time
amplitude or a divergent collective persistence time. But they do not exclude a stress-like source because the residual interaction force is generally correlated with the displacement field. We therefore next examine how the regular microscopic propulsion is reorganized by conservative interactions into spatially correlated interaction stress.}

\subsection {Emergent long-ranged interaction-stress correlations}
\label{sec:interaction-stress}

\SK{We now examine how active forcing is transmitted through
the conservative interaction-stress field. For an isotropic elastic solid, the stress and strain tensor can be written as,
\begin{equation}
 \sigma_{ij}^{\mathrm{el}}
 =
 \lambda_{\rm L}\delta_{ij}\epsilon_{kk}
 +2\mu\epsilon_{ij},
 \qquad
 \epsilon_{ij}
 =
 \frac{\partial_i u_j+\partial_j u_i}{2}.
 \label{eq:isotropicHooke}
\end{equation}
where, $\lambda_\mathrm{L}$ is first Lame coefficient and $\mu$ is shear modulus. Now, for a Fourier mode with wave vector $\vq$, the longitudinal and transverse projections gives
\begin{align}
 \sigma_{LL}^{\mathrm{el}}(\vq)
 &=
 \ii(\lambda_{\rm L}+2\mu)q\,u_L(\vq)
 =
 \ii M_Lq\,u_L(\vq),
 \nonumber\\
 \sigma_{TL}^{\mathrm{el}}(\vq)
 &=
 \ii\mu q\,u_T(\vq)
 =
 \ii M_Tq\,u_T(\vq),
 \label{eq:stressstrain}
\end{align}
where $M_L=\lambda_{\rm L}+2\mu$ and $M_T=\mu$.}

\SK{Defining the connected interaction-stress spectrum as
\begin{equation}
 S_{\mathrm{stress},\lambda}(q)
 =
 \avg{
 \left|
 \delta\sigma_{\lambda}^{\mathrm{int}}(\vq)
 \right|^2
 },
\end{equation}
the long-wavelength elastic constitutive relation gives
\begin{equation}
 S_{\mathrm{stress},\lambda}(q)
 =
 M_\lambda^2q^2\Cu_\lambda(q).
 \label{eq:stressCu}
\end{equation}
The elastic coefficients $M_\lambda$ are determined independently from the
direct mechanical response,
\begin{equation}
 K_{\mathrm{resp},\lambda}(q)
 =
 M_\lambda q^2.
\end{equation}
Consequently,
\begin{equation}
 \Cu_\lambda(q)\sim q^{-3}
 \quad\Longleftrightarrow\quad
 S_{\mathrm{stress},\lambda}(q)\sim q^{-1}.
 \label{eq:stressPrediction}
\end{equation}}

\SK{The Fourier transform of a $q^{-1}$ spectrum has the long-distance envelope
\begin{equation}
 G_{\mathrm{stress},\lambda}(r)
 \equiv
 \avg{
 \delta\sigma_\lambda^{\mathrm{int}}(\vrnew)
 \delta\sigma_\lambda^{\mathrm{int}}(\bm 0)
 }
 \sim r^{-(d-1)},
 \label{eq:realSpaceStressCorrelation}
\end{equation}
apart from the angular dependence associated with the tensor projection.
In two dimensions, this corresponds to an algebraic $1/r$ interaction-stress correlation.}

\SK{We calculate the microscopic interaction stress using the
Irving--Kirkwood expression in Eq.~\eqref{eq:IK}
\cite{IrvingKirkwood1950,Hardy1982,AdmalTadmor2010}. This construction contains
the LJ pair separation and the corresponding conservative pair force.
Figures~\ref{fig:stress}(a) and \ref{fig:stress}(b) show that both the transverse shear-stress and longitudinal normal-stress spectra increase
toward small $q$, approximately following the predicted $q^{-1}$ scaling. The transverse channel is particularly informative because $\sigma_{T}^{\mathrm{el}}(\vq)=\ii\mu q\,u_T(\vq)$ contains no compressional contribution and therefore directly demonstrates
long-ranged shear-stress correlations.}

\SK{The resulting microscopic picture is that active propulsion forces displace
the particles and deform the crystal. The conservative LJ interaction network
opposes these deformations and transmits them through correlated normal and
shear stresses. Hence, the observed $q^{-1}$ spectrum is an
activity-induced long-ranged correlation of the conservative interaction
stress, while the mechanical response itself remains conventionally elastic,
$K_{\mathrm{resp},\lambda}(q)\sim q^2$. Through the elastic stress--displacement
relation, these stress correlations are associated with the enhanced
displacement covariance $\Cu_\lambda(q)\sim q^{-3}$. The enhanced displacement fluctuations are mechanically carried by a long-ranged conservative interaction-stress network generated in the active steady state
\cite{Ronceray2015,Head2014}.}

\SK{To examine the temporal structure of the interaction-stress fluctuations, we
calculate the normalized connected autocorrelation function
\begin{equation}
 \widetilde C_{\mathrm{stress},\lambda}(q,t)
 =
 \frac{
 \avg{
 \delta\sigma_{\lambda}^{\mathrm{int}}(\vq,t_0+t)
 \delta\sigma_{\lambda}^{\mathrm{int}}(-\vq,t_0)
 }
 }{
 \avg{
 \left|
 \delta\sigma_{\lambda}^{\mathrm{int}}(\vq,t_0)
 \right|^2
 }
 },
 \label{eq:stressACF}
\end{equation}
where $\lambda=L,T$, the average is taken over time origins and independent realizations, and $\widetilde C_{\sigma,\lambda}(q,0)=1$ by construction. Figures~\ref{fig:stress}(c) show the transverse shear-stress correlation for $f_0=2.0$. Then we fit all the decaying correlations with $C_{\mathrm{stress},L}(q,t)/C_{\mathrm{stress},L}(q,0) \propto e^{\Gamma(q)t}$ and get the exponent $\Gamma(q,t)$ and plot it as a function of $q$ (see inset Figure~\ref{fig:stress}(c)). We see the exponent slows down as a function of $q^4$.
Next we measured the longitudinal normal-stress correlation (see Figure~\ref{fig:stress}(d)). In the passive crystal ($f_0=0$, inset), both correlations decay to zero on a very short time scale, indicating that the conservative stress fluctuations possess short-time memory. Activity strongly modifies this behaviour. For $f_0=2.0$, the long wavelength modes remain correlated for much longer times. This separation is more pronounced in the transverse modes, but is also present in the longitudinal modes. The observed increase of the relaxation time with decreasing $q$ demonstrates activity-induced long-wavelength slowing of the conservative interaction-stress field, basically activity retaining memory in the system, in close agreement with the observation made in \cite{sharma2025activity,goswami2025yielding}, where it was demonstrated that there is a close correspondence with the active driving and oscillatory shear response of the system.}

\begin{figure*}[t]
 \centering
 \begin{overpic}[width=0.48\linewidth]{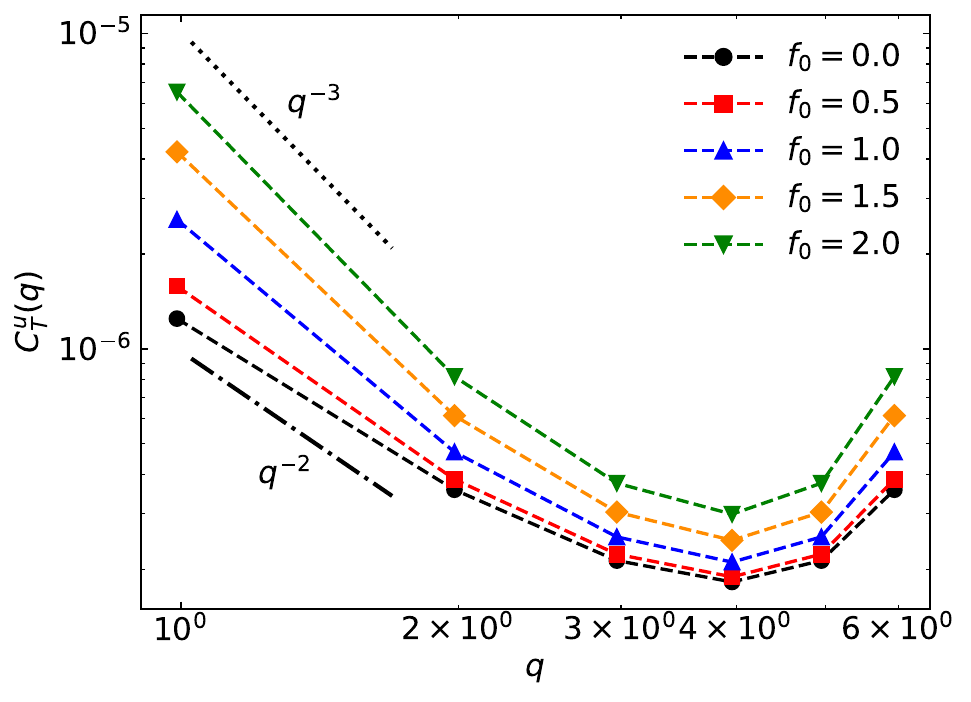}
    \put(18,14){\large\bfseries (a)}
\end{overpic}
 \begin{overpic}[width=0.48\linewidth]{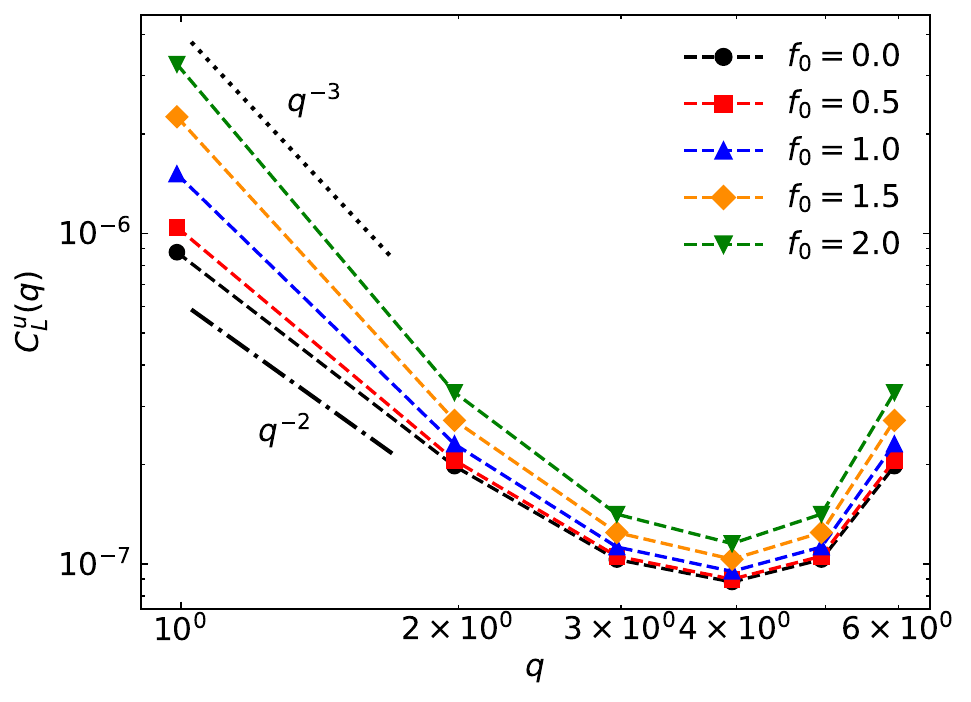}
    \put(18,14){\large\bfseries (b)}
\end{overpic} 
\begin{overpic}[width=0.48\linewidth]{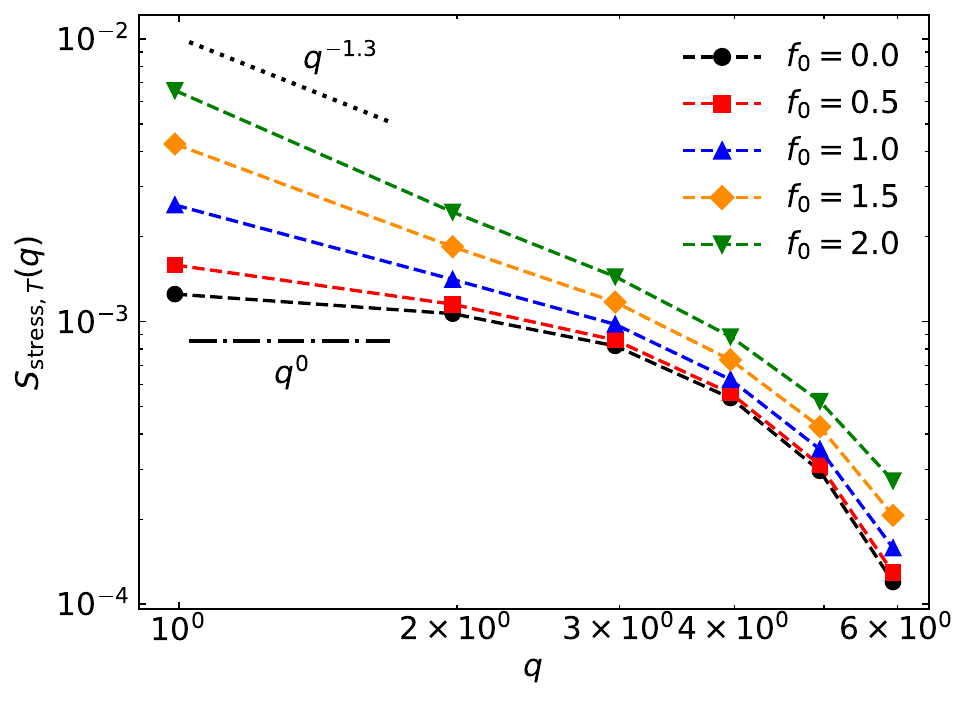}
    \put(18,14){\large\bfseries (c)}
\end{overpic} 
\begin{overpic}[width=0.48\linewidth]{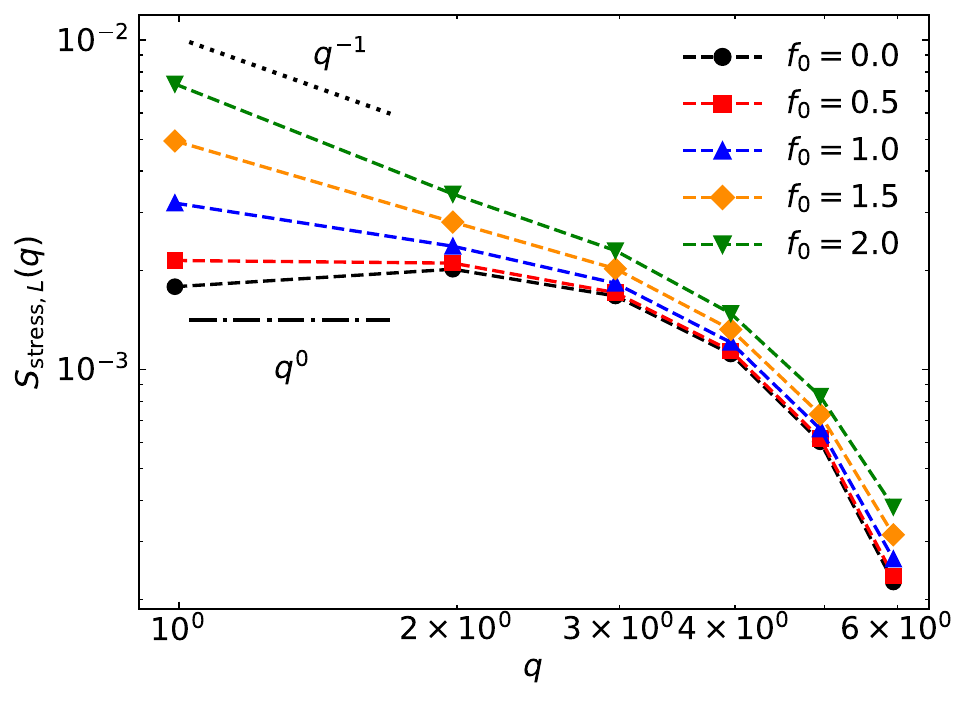}
    \put(18,14){\large\bfseries (d)}
\end{overpic}
 
\caption{\label{fig:3dSpectra}
Three-dimensional displacement and interaction-stress spectra of the active crystal.
(a,b) Transverse and longitudinal displacement covariances,
$\Cu_T(q)$ and $\Cu_L(q)$, respectively, for different propulsion strengths
$f_0$. The passive spectra are consistent with the elastic $q^{-2}$ behavior, whereas
increasing activity progressively enhances the lowest-$q$ modes.  For the
largest activity, $f_0=2.0$, both polarizations show a $q^{-3}$ divergence of displacement fluctuation at low $q$ limit. (c,d) Corresponding transverse and longitudinal interaction-stress spectra, $S_{\mathrm{stress},T}(q)$ and $S_{\mathrm{stress},L}(q)$, calculated from the conservative pair forces. The passive cases are nonsingular at small $q$, while activity produces a strong long-wavelength enhancement.  At $f_0=2.0$, the effective low-$q$ behaviors are approximately $S_{\mathrm{stress},T}\sim q^{-1.3}$ and
$S_{\mathrm{stress},L}\sim q^{-1}$.  The short black lines indicate the stated
power-law guides and are not independent fits.
}

\end{figure*}

\subsection{Three-dimensional consequences}
\label{sec:3d}

\SK{We next examine whether the activity-induced enhancement of displacement and interaction-stress fluctuations persists in a three-dimensional crystal. The translational equation of motion remains Eq.~\eqref{eq}, while the particle orientation now diffuses on the unit sphere. Its rotational dynamics can be written as
\begin{align}
 \dot{\bm n}_i&=\bm\eta_i^R\times\bm n_i,\nonumber\\
 \avg{\eta^R_{i\alpha}(t)\eta^R_{j\beta}(t')}
 &=2D_r\delta_{ij}\delta_{\alpha\beta}\delta(t-t'),
 \label{eq:3dOrientation}
\end{align}
where the stochastic rotation is interpreted in the Stratonovich sense, or
implemented by an exactly norm-preserving rotation, leads to
\begin{equation}
 \avg{\bm n_i(t)\cdot\bm n_i(0)}=e^{-2D_r|t|},
 \qquad \tau_p^{(3\mathrm D)}=\frac{1}{2D_r}.
 \label{eq:3dPersistence}
\end{equation}
We use an FCC reference crystal in a
cubic periodic box.  For $n_c$ conventional cells along each direction,
\begin{equation}
 N=4n_c^3,\qquad a=\left(\frac{4}{\rho}\right)^{1/3},
 \qquad d_{\mathrm{nn}}=\frac{a}{\sqrt{2}},                   \label{eq:FCC}
\end{equation}
where $a$ and $d_{\mathrm{nn}}$ are the cubic lattice constant and nearest-
neighbour separation.}

\SK{After removing the center-of-mass displacement in each frame, the Fourier mode is evaluated on the mean reference positions,
\begin{align}
 \delta\bm u(\vq,t)&=\frac{1}{\sqrt N}\sum_i
 \left[\bm u_i(t)-\frac{1}{N}\sum_j\bm u_j(t)\right]
 e^{-\ii\vq\cdot\bm R_i},\nonumber\\
 \vq&=\frac{2\pi}{L}(n_x,n_y,n_z).                          \label{eq:3dFourier}
\end{align}
Use of the instantaneous positions in the phase factor would mix the
displacement field with density fluctuations and is therefore avoided.}

\SK{For $\vq\ne0$, introduce the projectors
\begin{equation}
 P^L_{ij}=\hat q_i\hat q_j,\qquad
 P^T_{ij}=\delta_{ij}-\hat q_i\hat q_j.                      \label{eq:3dProjectors}
\end{equation}
There is one longitudinal polarization and a two-dimensional transverse
subspace.  We define the transverse spectrum per polarization,
\begin{align}
 \Cu_L(q)&=\avg{\delta u_i^*(\vq)P^L_{ij}\delta u_j(\vq)},\nonumber\\
 \Cu_T(q)&=\frac{1}{2}\avg{\delta u_i^*(\vq)P^T_{ij}
 \delta u_j(\vq)}.                                          \label{eq:3dCu}
\end{align}
Equivalently, for any orthonormal basis
$\{\hat{\bm e}_{T_1},\hat{\bm e}_{T_2}\}$ perpendicular to $\hat\vq$,
$\Cu_T=(\Cu_{T_1}+\Cu_{T_2})/2$.  This factor of $1/2$ is required for a
direct comparison with the single transverse branch in two dimensions.}

\SK{The three-dimensional Irving--Kirkwood tensor is still given by
Eq.~\eqref{eq:IK}.  A basis-independent projection is obtained from the
traction $t_i=\sigma^{\mathrm{int}}_{ij}\hat q_j$:
\begin{align}
 S_{\mathrm{stress},L}(q)&=\avg{|\hat q_i\delta t_i|^2},\nonumber\\
 S_{\mathrm{stress},T}(q)&=\frac{1}{2}\avg{\delta t_i^*P^T_{ij}\delta t_j}.
\label{eq:3dStressDef}
\end{align}
Thus the same normalization is used for transverse displacement and stress.}

\vskip +0.1in
\noindent{\bf Elastic relation in three dimensions:}
\SK{For an isotropic solid, Fourier transforming Hooke's law gives
\begin{equation}
 \sigma^{\mathrm{el}}_{ij}(\vq)=
 \ii\lambda_{\rm L}\delta_{ij}q_k u_k
 +\ii\mu(q_i u_j+q_j u_i).                                 \label{eq:3dHooke}
\end{equation}
For a longitudinal displacement $u_i=u_L\hat q_i$, projection of
Eq.~\eqref{eq:3dHooke} twice along $\hat\vq$ yields
\begin{equation}
 \sigma^{\mathrm{el}}_{LL}=\ii(\lambda_{\rm L}+2\mu)qu_L
 =\ii M_Lqu_L.                                               \label{eq:3dLongStress}
\end{equation}
For either transverse polarization $u_i=u_{T_a}e_{T_a,i}$, the dilation
$q_i u_i$ vanishes and
\begin{equation}
 \sigma^{\mathrm{el}}_{T_aL}=\ii\mu q u_{T_a},
 \qquad a=1,2.                                               \label{eq:3dTransStress}
\end{equation}
Taking variances and using the per-polarization definitions above gives
\begin{equation}
 S_{\mathrm{stress},L}=M_L^2q^2\Cu_L,\qquad
 S_{\mathrm{stress},T}=\mu^2q^2\Cu_T.                               \label{eq:3dStressCu}
\end{equation}
The two factors of $1/2$ in the transverse sector cancel. So Eq. (\ref{eq:3dStressCu}) gives the dimension independent exponent relation
\begin{equation}
 S_{\mathrm{stress},\lambda}(q)\sim q^{p_\lambda}
 \quad\Longleftrightarrow\quad
 \Cu_\lambda(q)\sim q^{p_\lambda-2}.                          \label{eq:3dExponentMap}
\end{equation}
Consequently, a nonsingular stress spectrum, $S_{\mathrm{stress},\lambda}\sim q^0$ produces the ordinary elastic covariance $\Cu_\lambda\sim q^{-2}$, whereas $S_{\mathrm{stress},\lambda}\sim q^{-1}$ produces the anomalous covariance $\Cu\sim q^{-3}$. In three dimensions, the corresponding real-space interaction-stress correlation decays as
\begin{equation}
G_{\mathrm{stress},\lambda}(r)\sim r^{-2},
\end{equation}
showing that the activity-induced stress correlations remain long-ranged.}

\vskip +0.1in
\noindent{\bf Three-dimensional spectra:}
\SK{Figure~\ref{fig:3dSpectra}(a,b) shows the displacement and Figure~\ref{fig:3dSpectra}(c,d) shows the corresponding interaction-stress spectra for transverse and longitudinal modes, respectively, obtained from the active FCC crystal.  In the common small-$q$ interval, both polarizations of $\Cu$ approach $q^{-3}$, while the corresponding traction spectra approach $q^{-1}$ rather than the passive $q^0$.  The paired trends satisfy Eq.~\eqref{eq:3dStressCu} and show that the same long-ranged normal and shear-stress structure occurs in three dimensions.}

\SK{Finally, the three-dimensional finite-size consequence follows directly from Eq.~\eqref{eq:W_def}.  With one longitudinal and two transverse branches,
\begin{align}
 W_3(L)&=\int_{2\pi/L}^{\Lambda}\frac{4\pi q^2\dd q}{(2\pi)^3 \rho}
 \left(\frac{A_L+A_{T_1}+A_{T_2}}{q^3}\right)\nonumber\\
 &=\frac{A_L+A_{T_1}+A_{T_2}}{2\pi^2 \rho}
 \ln\!\left(\frac{\Lambda L}{2\pi}\right)+O(1).             \label{eq:W3}
\end{align}
Therefore, $\Cu(q)\sim q^{-3}$ predicts a logarithmic growth of the
three-dimensional MSD plateau with system size, consistent with the behavior reported previously for three-dimensional active solids
\cite{Dey2025}. By contrast, the passive $q^{-2}$ covariance is infrared
convergent in three dimensions. Three dimensions are therefore marginal for
the anomalous active fluctuations: the MSD plateau grows logarithmically,
rather than algebraically, with the linear system size.}

\section{Discussion}

\SK{The covariance-inferred operator is cubic, yet all direct mechanical measurements are quadratic.  No contradiction arises because equilibrium FDT is absent: $\Hcov$ describes the stationary fluctuation distribution, whereas $K_{\mathrm{resp},\lambda}$ describes the force required to deform the solid.  Likewise, the measured $q^{-1}$ mode temperature is a real fluctuation--response ratio but not a standalone thermodynamic variable.}

\SK{As a side note, finite persistence time can produce a local slope $-3$ at the crossover between $q^{-2}$ and $q^{-4}$ over a small window of wave vector; thus, it remains a possibility that one is observing these anomalous displacement correlation simply due to finite size effects and asymptotically large system sizes (equivalent to very small wave vectors) one will see normal $q^{-2}$ behaviour of the displacement correlation variance, the same as the passive systems as proposed in \cite{Henkes2011, Kuroda2023}.  It cannot, however, explain an asymptotic $q^{-3}$ law at $q\to0$ unless an additional quantity becomes scale dependent.  The observed constant $\tau_f$, nonsingular $\mathcal D^f$, and local mobility exclude the previously proposed $\tau_f\sim q^{-1}$ route.  Varying $D_r$ and testing the predicted shift in the crossover wave vector might reveal a crossover to a true infrared regime.}

\SK{The microscopic-stress result adds information that covariance and response do not contain separately.  It shows where the collective distortion resides in the particle model: in correlated LJ bond forces, especially the transverse shear modes \cite{Takatori2014,Solon2015,Omar2020}.
The three-dimensional calculation sharpens the dimensional interpretation.
Once the two transverse modes are normalized per polarization, the elastic
mapping between stress and displacement is identical to the two-dimensional
one.  The simultaneous low-$q$ trends in Fig.~\ref{fig:3dSpectra} are therefore the expected static signature of the same mechanism, while the phase-space factor changes the integrated variance from $L$ to $\ln L$.}

\section{Conclusions}

\SK{The present results reveal some important clues on the previously
reported enhancement of Mermin--Wagner--Hohenberg fluctuations in active
crystals and glasses \cite{Dey2025}.  In the active Brownian crystal studied here, the displacement covariance changes from the passive
$\Cu_\lambda\sim q^{-2}$ form to an approximately $q^{-3}$ infrared spectrum. The earlier finite-size hierarchy and effective nonlinear dispersion are two consequences of this same covariance exponent.  In general, $\Cu\sim q^{-\psi}$ gives a marginal dimension $d_\ell=\psi$; hence $\psi=3$ predicts $W_2\sim L$, $W_3\sim\ln L$, and finite fluctuations in four dimensions.  Covariance inversion simultaneously gives
$\Hcov\sim|q|^3$ and $\omega_{\mathrm{cov}}\sim q^{3/2}$.}

\SK{One important result of the present work is that this covariance-defined dispersion is not the mechanical phonon dispersion.  Paired static forcing and frequency-dependent response both give $K_{\mathrm{resp},\lambda}\sim q^2$, so the response modes remain acoustic, $\omega_{0,\lambda}\sim q$.  The inverse covariance is therefore a statistical operator describing nonequilibrium mode occupation rather than a mechanically softened Hessian.  Correspondingly, $T_{\mathrm{mode},\lambda}=K_{\mathrm{resp},\lambda}\Cu_\lambda$ is strongly enhanced at small $q$ and has the inverse-$q$ behavior expected if the measured $q^2$ and $q^{-3}$ asymptotic forms overlap.}

\SK{The measurements also narrow the possible origin of this anomalous mode occupation.  The mobility is local within resolution, and neither the bare propulsion nor the reconstructed residual force develops the divergent persistence time or integrated intensity required by a simple additive-noise description. Thus, the present data exclude a cubic restoring kernel and the simplest singular mobility and forcing mechanisms, while a possibility of a crossover length scale that grows with persistence time remains.}

\SK{At the microscopic level, the longitudinal normal and transverse shear
components of the Irving-Kirkwood interaction stress develop approximately
$q^{-1}$ spectra. The relation $S_{\mathrm{stress},\lambda}\simeq M_\lambda^2q^2\Cu_\lambda$ connects these stress fluctuations directly to the anomalous displacement covariance. The paired response further shows that conservative interactions carry the ordinary $q^2$ restoring force, with negligible odd active-force feedback. The three-dimensional FCC calculation supports the same paired low-$q$ behavior, $\Cu_{L/T}\sim q^{-3}$ and $S_{\mathrm{stress},L/T}\sim q^{-1}$, and thereby provides a static microscopic counterpart to the previously observed logarithmic three-dimensional displacement growth. More broadly, the work demonstrates why covariance and response must be separated in active solids: their difference is not a technical detail but the statistical origin of enhanced Mermin-Wagner fluctuations beyond the equilibrium marginal dimension. The resulting physical picture is therefore precise: activity anomalously populates ordinary acoustic modes, and the conservative contact network transmits those collective distortions as long-ranged normal and shear stress. Instead of measuring MWH fluctuations using the covariance matrix, it would be more insightful to formulate a Hamiltonian that exhibits ABP-like motion. This would allow us to derive the Hessian directly and discover the origin of the enhanced displacement correlation \cite{bhattacharya2025thermostatting, bhattacharya2026perspective, bhattacharya2026active}.}

\SK{Finally, our understanding of stress fluctuations as the primary cause of anomalous displacement fluctuations, leading to enhanced long-wavelength density fluctuations and violation of the Mermin-Wagner-Hohenberg theorem in these solids, also corroborates recent findings of a close correspondence between active driving and the oscillatory shear response of these solids\cite{sharma2025activity, goswami2025yielding}. It suggests that shear modes carry information about the effect of active forces at the local level, especially at large length scales or small wave vectors. In the future, it will be interesting to see what kind of active forcing couples to the longitudinal modes and whether one can find the same signature in the corresponding longitudinal stress fluctuations. We believe that our results will help the community in the near future to develop a microscopic theory of active solids, keeping in mind that it is the stress fluctuations that get modified under active driving rather than an effective temperature.}

\begin{acknowledgments}
We acknowledge funding by intramural funds at TIFR Hyderabad from the Department of Atomic Energy (DAE) under Project Identification No. RTI 4007. SK wants to thank Hajime Tanaka, Chandan Dasgupta, and Sumilan Banerjee for useful discussions. SK acknowledges the Swarna Jayanti Fellowship grants DST/SJF/PSA01/2018-19 and SB/SFJ/2019-20/05 from the Science and Engineering Research Board (SERB) and Department of Science and Technology (DST). SK would like to acknowledge the research support from the MATRICES Grant MTR/2023/000079, funded by SERB and the J C Bose Grant ANRF/JBG/2026/000149/PS from the Anusandhan National Research Foundation (ANRF), Government of India. Some of the computations are done using the HPC clusters procured using Swarna Jayanti Fellowship grants and Core Research Grant CRG/2019/005373. SK also acknowledges the ANRF JC Bose grant ANRF/JBG/2026/000149/PS. 
\end{acknowledgments}

\bibliography{active_crystal_references}

\begin{thebibliography}{99}

\bibitem{Hohenberg1967}
P. C. Hohenberg, Existence of long-range order in one and two dimensions,
Phys. Rev. \textbf{158}, 383 (1967).

\bibitem{MerminWagner1966}
N. D. Mermin and H. Wagner, Absence of ferromagnetism or antiferromagnetism in
one- or two-dimensional isotropic Heisenberg models,
Phys. Rev. Lett. \textbf{17}, 1133 (1966).

\bibitem{Mermin1968}
N. D. Mermin, Crystalline order in two dimensions,
Phys. Rev. \textbf{176}, 250 (1968).

\bibitem{Flenner2015}
E. Flenner and G. Szamel, Fundamental differences between glassy dynamics in
two and three dimensions, Nat. Commun. \textbf{6}, 7392 (2015).

\bibitem{Vivek2017}
S. Vivek, C. P. Kelleher, P. M. Chaikin, and E. R. Weeks,
Long-wavelength fluctuations and the glass transition in two dimensions and
three dimensions, Proc. Natl. Acad. Sci. USA \textbf{114}, 1850 (2017).

\bibitem{Illing2017}
B. Illing, S. Fritschi, H. Kaiser, C. L. Klix, G. Maret, and P. Keim,
Mermin--Wagner fluctuations in 2D amorphous solids,
Proc. Natl. Acad. Sci. USA \textbf{114}, 1856 (2017).

\bibitem{Li2019}
Y.-W. Li \emph{et al.}, Long-wavelength fluctuations and anomalous dynamics in
two-dimensional liquids, Proc. Natl. Acad. Sci. USA \textbf{116}, 22977 (2019).

\bibitem{Marchetti2013}
M. C. Marchetti \emph{et al.}, Hydrodynamics of soft active matter,
Rev. Mod. Phys. \textbf{85}, 1143 (2013).

\bibitem{Fodor2016}
\'E. Fodor, C. Nardini, M. E. Cates, J. Tailleur, P. Visco, and F. van Wijland,
How far from equilibrium is active matter?,
Phys. Rev. Lett. \textbf{117}, 038103 (2016).

\bibitem{Takatori2014}
S. C. Takatori, W. Yan, and J. F. Brady,
Swim pressure: Stress generation in active matter,
Phys. Rev. Lett. \textbf{113}, 028103 (2014).

\bibitem{Omar2020}
A. K. Omar, Z.-G. Wang, and J. F. Brady,
Microscopic origins of the swim pressure and the anomalous surface tension of
active matter,
Phys. Rev. E \textbf{101}, 012604 (2020).

\bibitem{Dey2025}
S. Dey, A. Bhattacharya, and S. Karmakar,
Enhanced long wavelength Mermin--Wagner--Hohenberg fluctuations in active
crystals and glasses, Nat. Commun. \textbf{16}, 5498 (2025).

\bibitem{Henkes2012}
S. Henkes, C. Brito, and O. Dauchot,
Extracting vibrational modes from fluctuations: A pedagogical discussion,
Soft Matter \textbf{8}, 6092 (2012).

\bibitem{Kuroda2023}
Y. Kuroda, H. Matsuyama, T. Kawasaki, and K. Miyazaki,
Anomalous fluctuations in the homogeneous fluid phase of active Brownian
particles, Phys. Rev. Research \textbf{5}, 013077 (2023).

\bibitem{IrvingKirkwood1950}
J. H. Irving and J. G. Kirkwood,
The statistical mechanical theory of transport processes. IV. The equations of
hydrodynamics, J. Chem. Phys. \textbf{18}, 817 (1950).

\end{thebibliography}

\end{document}